\documentclass{statsoc}
\usepackage[a4paper]{geometry}
\usepackage{graphicx}
\usepackage{graphics}
\usepackage[textwidth=8em,textsize=small]{todonotes}
\usepackage{amsmath}
\usepackage{natbib}
\usepackage{amsmath}
\usepackage{natbib}
\usepackage{epsfig}
\usepackage{latexsym}
\usepackage{color}
\usepackage{mathrsfs}
\usepackage{appendix}
\usepackage[figuresright]{rotating}
\usepackage{dsfont}
\usepackage{longtable}
\usepackage{amsmath,amssymb,amsfonts, eurosym}
\usepackage{psfrag}
\usepackage{enumerate}
\RequirePackage[caption=false]{subfig}
\usepackage{url}
\usepackage{comment}
\usepackage{pdfsync}
\usepackage{colortbl}
\usepackage{epstopdf}
\usepackage[normalem]{ulem}
\usepackage{multirow}
\usepackage[utf8]{inputenc}
\usepackage[T1]{fontenc}
\usepackage{hyperref}
\hypersetup{
    colorlinks=true,
    linkcolor=black,
    citecolor=black,
    filecolor=black,
    urlcolor=black,
}

\DeclareMathOperator*{\cls}{cls}
\DeclareMathOperator*{\ilr}{ilr}
\DeclareMathOperator*{\clr}{clr}

\DeclareMathOperator{\pen}{pen}
\DeclareMathOperator{\T}{^{\top}}

\def\1{\mathds{1}}
\newcommand{\e}{\mathds{E}}
\newcommand{\var}{\mathds{V}\text{ar}}
\newcommand{\cov}{\mathds{C}\text{ov}}

\graphicspath{{figures/}}

\title[Generalised functional additive mixed models with compositional covariates]{Generalised functional additive mixed models with compositional covariates for areal  Covid-19 incidence curves}

\author[Matthias Eckardt {\it et al.}]{Matthias Eckardt}
\address{Chair of Statistics, Humboldt-Universit{\"a}t zu Berlin, Berlin, Germany.}
\email{m.eckardt@hu-berlin.de}

\author[Matthias Eckardt {\it et al.}]{Jorge Mateu}
\address{Department of Mathematics, University Jaume I, Castell\'{o}n, Spain.}

\author[Matthias Eckardt {\it et al.}]{Sonja Greven}
\address{Chair of Statistics, Humboldt-Universit{\"a}t zu Berlin, Berlin, Germany.}

\begin{document}
\maketitle

\begin{abstract}
We extend the generalised functional additive mixed model to include (functional) compositional covariates carrying relative information of a whole. Relying on the isometric isomorphism of the Bayes Hilbert space of probability densities with a subspace of the $L^2$, we include functional compositions  as transformed functional covariates with   constrained  effect function. The extended  model allows for the estimation of linear, nonlinear and  time-varying effects of scalar and functional covariates, as well as (correlated) functional random effects, in addition to the compositional effects. We use the model to estimate the effect of the age,
sex and smoking (functional) composition of the population on regional Covid-19 incidence data for Spain, while accounting for  climatological and socio-demographic covariate effects and spatial correlation.  
\end{abstract}

\keywords{Covid-19, compositional data analysis, functional compositions, functional data analysis, functional regression, function-on-function regression}

\section{Introduction}

Understanding the infectious disease dynamics of the Covid-19 (Coronavirus disease 2019) pandemic and its potential associations with  different exogenous environmental, socio-economic and health-related variables  has become an important  challenge of current interdisciplinary research. Although massive data is collected,  the long-term interplay of the local numbers of daily Covid-19 cases with sets of different (potentially time-varying) risk factors still remains an open and challenging topic. In particular, this includes the  effects of the composition  of the local population (age, sex, smoking etc.) on the spread of the disease, which has not been investigated to the  best of our knowledge. This paper aims to fill this gap by  extending the generalised functional additive mixed model (GFAMM) of \cite{FGAMM} to the case where parts of the covariate set are finite or infinite compositions, i.e.\ multivariate or functional covariates carrying relative information of a whole. The proposed formulation  allows to model the local Covid-19 dynamics conditional on various types of exogenous variables,  including among others the population density (a scalar), the average temperature (a function of time), the smoking status (a finite composition), the age composition (a functional composition or infinite composition or density) and the regional  structure (a grouping factor with spatial correlation).
To this end, using areal Covid-19 incidence data collected   in Spain daily until the vaccination onset, the  local incidence curves are modelled taking a functional data analysis \citep{ramsey2005} perspective. 

In functional data analysis, the response curves are considered as realisations of some stochastic process with continuous support, such as time. While the process itself could theoretically be observed for  any point at arbitrary resolutions, the curves are only measured on a discrete grid. A suitable class of functional data analysis techniques for the present purpose are functional regression models which can, in general,
be summarised into  scalar-on-function (where only the covariates are functions), function-on-scalar (where response curves are related to scalar covariates), and function-on-function models (where both the response and the regressor are functions). See \cite{Morris2015} and  
\cite{doi:10.1177/1471082X16681317} for a general review of different functional regression specifications. Regressions for non-Gaussian functional responses (e.g.\ counts) include e.g. the generalised function-on-scalar model \citep{https://doi.org/10.1111/biom.12278} and also the GFAMM \citep{FGAMM}, which provides a flexible regression framework for possibly non-Gaussian functional responses on potentially irregular or sparse grids using basis function representations of the fixed and/or random  effects of scalar and/or functional covariates. 
We note that a complementary approach for functional regression models observed on equidistant grids is presented in  
\cite{10.2307/3647565} and subsequent work, 
which uses a Bayesian estimation approach based on wavelet transformations of the functions that is, however, not suited to  generalised functional responses. A general comparison of both the basis function and the wavelet transformation approaches is given by  \cite{doi:10.1177/1471082X16681875} and \cite{rejoinder}.

Although different (generalised) functional regression specifications exist, only a very small body of the literature discusses extensions of functional regression  models to the case where the responses or parts of the covariate set are finite or infinite compositions. Predominantly, the literature focused on extensions to density-valued (functional compositional) responses while extensions to density-valued covariates appeared only rarely (see \cite{PETERSEN2021} for a recent review of different statistical approaches to density-valued quantities).  Apart from a Hilbertian random variable treatment of density-valued explanatory variables \citep{SIERRA20151192} or density-valued explanatory variables and responses  \citep{PARK2012397} embedded into the standard $L^2$ space, density-valued responses were covered in the additive regression model of \cite{doi:10.1080/01621459.2019.1604365}, where the outcome is mapped to the space of unrestricted square integrable functions in a pre-processing step, and also the nonlinear space formulation of \cite{10.1214/17-AOS1624}. However, while the $L^2$ space approach does not account for the constrained nature of the response, the additive regression and the nonlinear space formulations are affected by instabilities of the pre-transformations \cite[see][]{happ2019general} and do not allow for a straightforward interpretation of the regression coefficients, respectively. Different from the above   approaches,  \cite{RePEc:eti:dpaper:17015, TALSKA201866} and \cite{maier2021} applied a  Bayes Hilbert space formulation 
\citep{doi:10.1111/anzs.12074}  to include  density-valued responses into a functional regression framework where the estimation uses the centred log ratio (clr) transformation to map the response to a subspace of the $L^2$ with integration-to-zero constraint. 
Although this approach provides a promising framework for density-valued response regression, extensions which (additionally) allow for density-valued explanatory variables remain extremely limited. A first  linear Bayes Hilbert space regression model for scalar responses and density-valued covariates was proposed by \cite{Talska2021} using a constrained spline representation 
\citep{Machalova2021} 
 and extended further by \cite{DoD2021:MOX}, allowing for both density-valued responses and   covariates. While this model allows for linear effects of the  functional composition on the response, extensions to generalised (functional) additive models where parts of the predictor  are finite or infinite compositions still remains an open topic. We fill this gap within the flexible GFAMM framework. We build on
 \cite{doi:10.1111/rssc.12283}, who presented an extended  predictor for the generalised additive model for scalar responses with compositional covariates, by  extending this idea to functional responses and functional compositions as covariates. In particular, 
 both finite and infinite compositions are included into a general   structured predictor through suitable basis function representations that account for the  constrained covariate nature. 

The remainder of this paper is structured as follows. Section \ref{sec:data} briefly reviews the history of the pandemic in Spain and recent findings on potential risk factors for the spread of the disease that inform our selection of covariate effects. It also  provides information on the data sources and (generated) variables used in the analysis, and presents a descriptive analysis of the different covariates included in the GFAMM. An introduction to this model is presented in Section \ref{sec:model}. In particular, extensions of the covariate set to finite and infinite compositions are discussed in Section \ref{sec:coda}. An application of the proposed extension to Spanish Covid-19 incidence data is given in Section \ref{sec:results}. The paper concludes with a discussion in Section \ref{sec:final}.

\section{Covid-19 data for Spain}\label{sec:data}


\subsection{A small history of the Covid-19 pandemic in Spain}
Within  three months of the official notification of a small regional outbreak 
in Wuhan, China, in late December 2019, Spain was facing one of the highest infection and, in particular, mortality rates among the European countries \citep{Covid:spain}. Within four weeks of the first tourist-based case on the Island of La Gomera in  January  and the first domestic hospitalisation  on 15th February 2020,   Spain witnessed a large but spatially strongly heterogeneous increase in numbers of Covid-19 infections with a clear concentration in large metropolitan conurbations 
\citep{HENRIQUEZ2020560}. These marked regional differences and early local peaks in e.g. Madrid, may in part be explained by the regional mobility from and to the Spanish capital \citep[cf.][]{Mazzoli2020.05.09.20096339}. 
Using data for Catalonia, \citet{ComaRedone039369} provided some evidence that  early local Covid-19 cases may have been masked by excess of influenza notifications  between 4th February 2020 and 20th March 2020 as polymerase chain reaction (PCR) tests were restricted to hospital-admitted patients only and general practitioners were asked to diagnose Covid-19 infections without PCR   confirmation. Due to this uncertainty, early  cases and deaths in private and nursing homes may have been excluded from the official reports for this period. However, previously undetected symptom-based cases and deaths were subsequently added to the official notification system such that we can plausibly assume that any remaining bias can be neglected in the present study. 

In response to the rapid increase in mortality, particularly among the elderly (potentially multi-morbid) parts of the population, and the transmission dynamics of the disease, the Spanish National Government imposed a series of global regulatory interventions to suppress the spread of the virus. A first strict lockdown excluding only essential services (e.g. food, health) and some economic subsectors was imposed on 14th March. This measure was tightened in subsequent actions by placing strict entry refusal at the Spanish borders on 17th March and prohibiting any non-essential activities within the period from 30th March to 12th April 2020. In consequence, these  restrictions yielded a dramatic reduction in overall regional mobility. Ended on 21st June, this lockdown remains the only global action of the Spanish Government against the Covid-19 pandemic. Although facing severe increases in numbers of Covid-19 cases in the second half of 2020 and early 2021, the public health responsibilities were delegated back to the local governments of the autonomous communities by 21st June and  only local but no further global restrictions  were reimposed. These (mostly soft) local measures, however, show a strong heterogeneity, and information on the exact timing, nature and extend of the different imposed restrictions was not available  
from official sources for our study. 

\subsection{Recent findings on risk factors for the spread of Covid-19}

Apart from  a higher risk of Covid-19 infections caused by the increase in immunosenescence for the older ages \citep{10.1093/ppar/praa023}, a clear association of higher age with the 
development of severe symptomatic Covid-19 infections,  hospitalisation and  fatality rates is stressed in the literature \citep[see e.g.][]{Tiruneh2021}. Besides  certain health conditions and comorbidities \citep{Du2021}, \citet{Wolff2021} identified smoking as an important co-factor. In particular, active smokers or non-active smokers with a clear smoking history face 
an increased risk for the development of symptomatic Covid-19 \citep{Hopkinsonthoraxjnl,Guelsen2020}.
Apart from these findings, e.g.  \cite{https://doi.org/10.1002/hpm.3189} reported a close relation between densely populated regions and contact rates between different (potentially infected) individuals, which positively impact the disease transmission and, in turn, the reproduction rate of the disease. This idea is also supported by \cite{paez2021} who reported a clear positive effect of mass transport systems on the incidence. These authors also reported a positive association of the disease with wealthier regions, i.e.\ regions with a higher GDP per capita, with a potential explanation via a connection between wealth and the regional level of  globalisation, i.e.\ international trade and travel.   

The effect of climatological and environmental covariates on the spread of the disease is less consistent. In line with results on similar pathogens, which suggest that the virus is more stable and transferable in conditions of low temperature and low humidity, \cite{paez2021} found a negative effect of higher values in  temperature and humidity on the incidence of the disease, contrasted with a positive impact of sunshine. In a systematic review,  \cite{10.1371/journal.pone.0238339} found a positive effect of cold and dry weather conditions on the seasonal viability and transmissibility of Covid-19.  \citet{Takagi2020.05.09.20096321} reported an inverse association of temperature, air pressure, and ultraviolet light with the prevalence of the disease, while 
\citet{HOSSAIN} draw mixed conclusions based on both positive and negative effects of the weather characteristics.  \citet{Shahzad2020} highlighted a clear positive effect of bad air quality on the transmission of Covid-19 whereas temperature serves only as a contributory factor, with higher temperatures reducing the spread of the disease. Summarising the findings of 23 articles in a systematic review, \citet{ijerph18020396} found a clear association of temperature and Covid-19 and also a significant association of humidity and Covid-19 which, however, was derived from mixed results. Using a nonlinear effect specification, \cite{WU2020139051} found a negative association of high temperature and also high humidity with the daily number of Covid-19 cases and associated deaths.

\subsection{Data}
\subsubsection{Data sources and variables}

The data was compiled from different sources, most commonly information provided by the regional governments.  It originates from a collaborative data project by the geovoluntarios community  (\url{https://www.geovoluntarios.org}), Centro de Datos Covid-19 and ESRI Spain and provides information on the daily numbers of Covid-19 cases for 52 Spanish provinces, each of which subsumes numerous local administrative units. It covers the period from 5th January 2020 to 19th January 2021 until just before vaccinations  became more widespread. Covid-19 cases are defined as probable infections without test information  or confirmed infections based on positive test results derived from (i)  PCR, antibody, and  antigen detection or Elisa techniques, and (ii) reported by other laboratories - showing a clear majority of results derived from positive PCR tests. In contrast, notifications based on antibody tests (with less precise timing information on the time of infection) contributed only at rather small and also spatially-varying rate. Restricted to the first Spanish Covid-19 wave only, the highest regional proportion of antibody  based test  results relative to all cases reported appeared for the provinces of Cuenca (5.06\%) and Albacete (2.34\%). We thus use the complete incidence counts based on all tests.
To account for a potential delay between the date of the test and the notification date caused by the individual testing procedures, the dates were shifted back by the provider using a three days lag. Different from  data used in this study, official periodical data releases on the Covid-19 pandemic through the Spanish National Government cover only  information at a coarser level of spatial aggregation, i.e.\  18  so-called \textit{autonomous communities}, to which the 52 Spanish provinces belong. Note that 2.05 \% (resp. 0.02 \%) of the Spanish population received the first (resp. second) vaccination before 19th January 2021. Due to this relatively small proportion of partly vaccinated inhabitants, any potential confounding effects of the vaccination action on the incidence data is assumed to be ignorable.

To investigate (potential time-varying) effects on the spread of the pandemic over space and time, we linked this data to climatological,  socio-economic and demographic  information, recorded for the  provinces and time period under study (see Table 1 in the Supplement for a detailed description of the variables). Daily climatological information including the average daily temperature (in $^\circ C$), humidity (in \%), maximal wind speed (in km/h), sun hours (in h) and precipitation (in mm) were collected for each province from the state meteorology agency (AEMET) and the ministry of agriculture, fisheries, and food (MAPA). The original data exhibited some missing values, in particular $0.5\%$ for average temperature,  $0.77\%$ for maximal wind speed, $ 2.07\%$  for sun hours, $5.34\%$  for precipitation, and $0.5\%$  for humidity - most likely  caused by transmission or technical problems. Any of these missing values were imputed by linear interpolation using the \texttt{imputeTS} package  \citep{imputeTS} in R. As  information on the daily solar exposure was not provided at all for Malaga, missing values for sun hours for Malaga were replaced by average values computed from the neighbouring provinces, i.e.\ Cadiz and Granada.  For the precipitation variable, the reported values show a large number of zeros in the daily amount of precipitation at the province level (ranging from $36$ days at minimum to $215$ days at maximum) as well as skewness with extreme peaks of $150mm$. For this reason, we summarised the original information into a binary variable indicating the absence or presence of rain per province on a daily basis. In addition to this rain indicator, we computed the log transformation of the nonzero precipitations. Next, all weather information was shifted using a 5-days lag to account for the time lag between infection and symptom onset, as we want to investigate weather effects on the infection probability and assume symptom onset to be strongly correlated with the timing of the performed  Covid-19 test. While the 5-day lag is supported by the findings of  e.g. 
\cite{jcm9020538}  who reported an average incubation period (defined as the time from the infection to symptom onset)  by around 5  days, we tested for the effect of different lag specifications in a sensitivity analysis.  

Regional  socio-demographic information on the number of inhabitants, the gross domestic product (GDP) per capita, the proportion of males, and age pyramids (0-100 years)  were collected from public data provided by the national statistics institute. This source was also used to compute the smoking composition of the population 
(categorised in daily, occasional, ex- and non-smokers) at the provided coarser level of the \textit{autonomous communities}, which we then assigned to 
all provinces within this community. 

\subsubsection{Generated variables}
In addition to the above data, we generated different  variables to control for the regional and geographical characteristics of the individual spatial units. First, to account for a potential impact of public mass transportation systems on the transmission and spread of the virus, we generated a binary variable indicating whether or not a province offers access to  a metro or subway system. 
In addition, we generated a second binary variable  indicating whether or not a province offers direct access to the Mediterranean Sea or the Atlantic Ocean. 
Besides a  higher population density in the coastal regions compared to the inland provinces (except for the metropolitan regions), the coastline is commonly strongly affected by high numbers of incoming tourists during the summer and public vacation periods, which might  serve as an acceleration factor for the risk of infection. Finally, to account for (i) potential temporal variation in the regional notification systems, and (ii) the effect of the global lockdown measures, we generated six weekday dummy variables treating Sundays as reference and three global lockdown dummies. Each of these lockdown indicators corresponds to one of the three subsequent global lockdown periods with different measures  imposed  during the first wave in 2020, i.e.\ (i) May, 11th to May, 24th, (ii) May, 25th to June, 7th, and (iii) June, 8th to June, 21st.

\subsection{Data description}\label{sec:datadesc} 
\begin{figure}[h!]
 \centering
\makebox{
    \includegraphics[scale=0.5553]{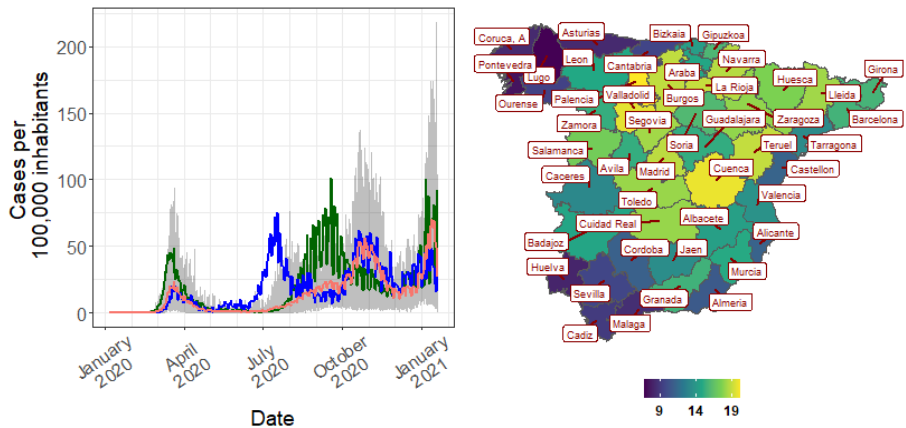}}
    \caption{Distribution of daily Covid-19 cases per 100,000 inhabitants over time and space: (a) mean (red) and regional incidence for all provinces (grey), as well as for Madrid (green) and Lleida (blue)   over 381 days,  and (b) average incidence over 381 days at province level, restricted to continental Spain.}
        \label{fig:outcome_means}
\end{figure}

The response curves show strong regional heterogeneity, with the highest numbers of cases in the provinces of Madrid and Barcelona. The highest peak ($y=6750$) appeared for Madrid on  September 18th, 2020, contrasted with e.g.\ only $y=77$ on that date recorded for the province of Lleida. To better understand the similarities among the spatial disease patterns, we calculated the regional incidence rates per 100,000 inhabitants, and its average version, computed as the mean rate per region over all 381 days (see Figure \ref{fig:outcome_means}). The incidence curves (left panel) reflect a clear positive deviation for Madrid (green) and also Lleida (blue) from the mean curve (red) for the period from June to October 2020 - with a  temporal delay in the   second wave for Madrid compared to Lleida. The corresponding regional averages (see right panel) indicate a clear spatial pattern, with Palencia and Cuenca showing  the highest average incidence rates with 20.5 and 19.66 cases per 100,000 inhabitants, respectively. These high average values are contrasted with relatively small reported rates  for e.g.  Lugo (6.77). The clustering of small and high average rates suggests a positive spatial autocorrelation structure in the data, which is supported by highly significant results for Moran's $I$ \citep{moran}  index (restricted to continental provinces). See Figures 1 and 2 in the Supplement for the spatial distribution over all Spanish provinces and communities including the Spanish islands and African enclaves.    

\begin{figure}[htbp!]
 \centering
 \makebox{
    \includegraphics[scale=.44]{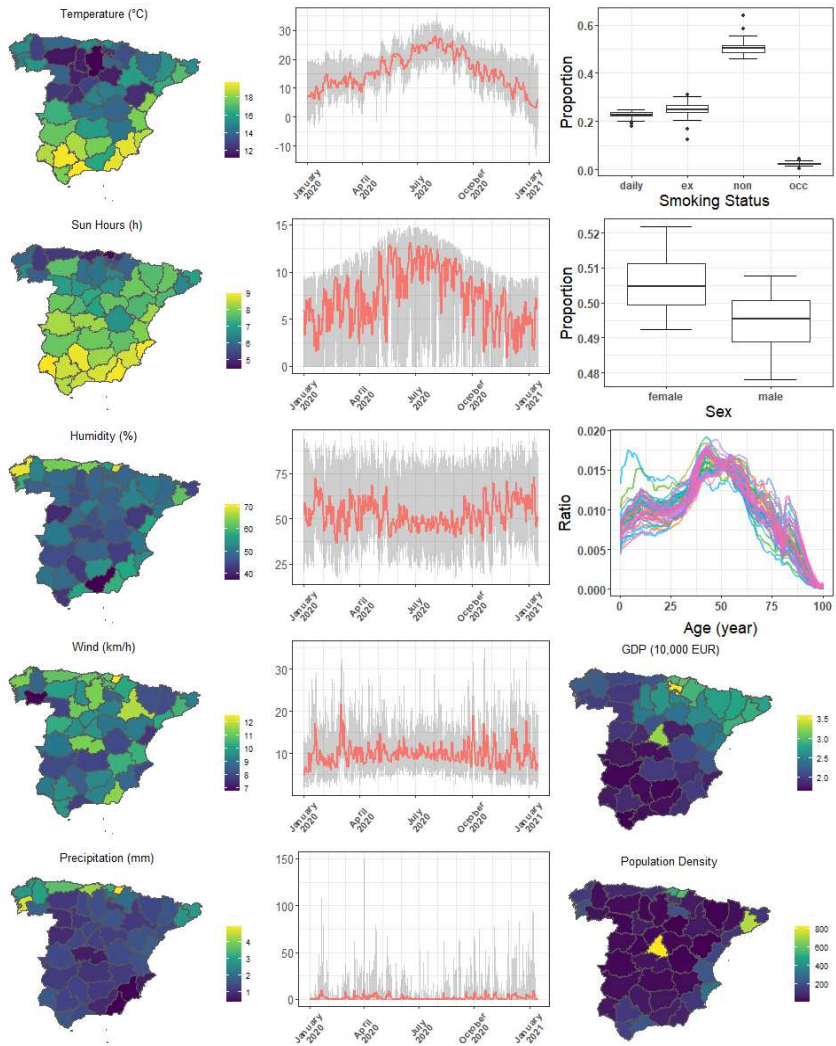}}
    \caption{Distributional characteristics of the climatological, compositional and socio-demographic  covariates:  temporally averaged spatial variation (left column), daily variation over time (central column, mean in red) of regional weather characteristics at province level, and regional variation of compositional and scalar covariates (right column).}
  \label{fig:temp_raw}
\end{figure}

Figure \ref{fig:temp_raw} illustrates the patterns of the scalar, functional and compositional covariates. Both the averaged spatial patterns over time of mean temperature and solar exposure (left column of this plot) show a general increase   from the northern to the southern parts of Spain. The temporal means for humidity and precipitation reflect some spatial heterogeneity with higher values computed for the northern areas contrasted with lower values for both variables at the Mediterranean coastline. Different from the other four weather variables, the maximum wind speed exhibits little spatial correlation, with the highest values reported for Gipuzkoa in the north. Over time (central column), both average temperature and sun hours  show high values during the summer contrasted with low values in the winter.  At the same time,  humidity, precipitation and  maximum wind speed reflect less clear temporal patterns. Although some variation and higher peaks are shown for humidity and wind speed in winter, spring and autumn compared to the summer period, the mean precipitation levels remain constantly at low values over time. 

For the compositional covariates depicted in the right column of this plot we found a clear dominance of  non-smokers over daily and ex-smokers in all 52 provinces, with occasional smokers (occ) constituting the smallest part of the regional populations. For the sex composition,  a small  dominance of females over males exists in all provinces.  
The normalised age curves computed from the age pyramids  show a clear mode, with the largest population mass around 50 years. A second smaller mode and large variation can be seen for younger ages of around 10 years, while the densities decrease roughly monotonically and consistently across provinces for ages older than 55 years.          
Finally, the spatial patterns for the socio-demographic time-constant variables, shown in the bottom two right panels of Figure \ref{fig:temp_raw} reflect a clear spatial variation of the individual GDP (in 10,000 Euro), with lower values in the south contrasted with higher values in the northern provinces of Spain and also Madrid. The population density shows a strong heterogeneity with the highest values  for the metropolitan provinces of Madrid and Barcelona, but also the provinces of Bizkaia and Gipuzkoa.

\section{Model specification}\label{sec:model}

\subsection{The generalised functional additive mixed model}

We extend the GFAMM specification of \cite{FGAMM} to include compositional and functional compositional covariates, such as for gender, smoking status and age composition in the present context. We adopt a general representation in the form of a structured additive regression model with $Y_i(t)\sim\mathcal{F}(\mu_i(t))$,  where $Y_{i}(t)$ in our setting is the 
number of Covid-19 cases  in province $~i=1,\ldots, 52$ at time $t \in \mathcal{T}$. 
In general, we assume that $Y_i(t)$ pointwise follows a (here count) distribution $\mathcal{F}$ with conditional expectation $\e\left[Y_i(t)\right|x_{it},t]=\mu_i(t)$ and is recorded over a domain $\mathcal{T}$, here covering the 381 days of observations. To account for potential overdispersion of the response, we here assume a  quasi-Poisson model for the Covid-19 incidences such that the variance is related to the mean through the overdispersion parameter $\xi$, i.e.\ $\var\left[Y_i(t)|x_{it},t\right]=\mu_i(t)\xi$. 
To achieve high flexibility  of the model,  the mean $\mu_i(t)$ is related to an additive predictor $\eta_i(t)$ through a known link function $g$, 
\[
g(\mu_i(t)) = \eta_i(t) = \sum_{r=1}^R f_r(x_{rit},t).
\]
Each additive term $f_r(x_{rit},t)$ is a smooth function of the argument $t$ of the outcome - also implying smoothness of the response mean $\mu_i(t)$ - and of a  subset $x_{rit}$ of the complete covariate set $x_{it}$. 

The above formulation allows for linear, nonlinear and time-varying effects of grouping factors, functional and potentially time-varying scalar   covariates,  as well as functional random effects, where the form of $f_r(x_{rit},t)$ is determined by the covariates in $x_{rit}$. For example, for  a functional intercept that varies over $t$, $f_r(x_{rit},t)$ simplifies to $\beta_0(t)$. For a  smooth effect of a scalar $x_{ir}$ that is constant over $t$, $f_r(x_{rit},t)$  becomes $f_r(x_{ir})$ (and $f_r(x_{ir},t)$ in the time-varying case), whereas linear effects of $x_{ir}$ that vary over $t$ correspond to $x_{ir}\beta(t)$. Linear time-varying effects of a functional covariate $x_{ir}(s), s \in \mathcal{S}$,  are included as $\int_{\mathcal{S}}x_{ir}(s)\beta(s,t)ds$, while a concurrent effect for $\mathcal{S} = \mathcal{T}$ can be included as $f_r(x_{ir}(t),t)$ or $f_r(x_{ir}(t))$. For a grouping variable $c$ with $M$ levels, scalar and functional random effects $\gamma_c$ and $\gamma_c(t)$ are included as zero mean Gaussian variables with a potentially general specified correlation structure, and as Gaussian processes $\mathcal{T} \times \lbrace 1,\ldots M\rbrace \rightarrow \mathds{R}$ with general covariance function $\cov(\gamma_c(t), \gamma_{c^{\prime}}(t^{\prime}))$  that is smooth in $t, t^{\prime}$. 
This specification also allows to control for the  spatial correlation of the different levels of $c=i$ (formalised through the  precision matrix of a Markov Random Field (MRF) with known correlation based  on the  planar neighbourhood structure) in the construction of (potentially spatially correlated) smooth residual curves for the individual locations. In the case of  multiple random effects, a mutual independence assumption is placed between the individual random terms. See \cite{Scheipl:FAMM, FGAMM} for a full detailed list of potential covariate specifications. 

However, compositional and functional compositional covariates have not yet been included into this framework, as occur in our context due to the available  information on the age-, sex- and smoker-compositions of the provinces. To include both kinds of effects into the GFAMM framework, we extend the predictor $\eta_i(t)$ by  $\eta_i^{comp}(t)$  summarising the effects of the compositional and functional compositional covariates. We note that this  approach is similar in spirit to \cite{doi:10.1111/rssc.12283}, who introduced an extended additive predictor to incorporate the effect of a compositional covariate in the context of generalised additive models for scalar responses. We extend this approach to functional responses and functional compositional covariates and give details in Section \ref{sec:coda} below.  

\subsection{Basis representation and estimation}
Having  $n$  functional observations $y_i(t)$ on a grid  of $T_i$ points  $\mathbf{t}_i=({t}_{i1},\ldots,{t}_{iT_i})\T$ available, the model can be fitted through a penalised (quasi-)likelihood approach based on the $(\sum_{i=1}^n T_i)$-vectors $\mathbf{y}=(\mathbf{y}_1\T,\ldots,\mathbf{y}_n\T)\T$ and  $\mathbf{t}=(\mathbf{t}_1\T,\ldots,\mathbf{t}_n\T)\T$ of the concatenated response curves and their arguments, respectively, where  $\mathbf{y}_{i}=(y_{i1},\ldots,y_{iT_i})$ with  $y_{il}=y_i(t_l), l=1,\ldots, T_i$. Here, $T_i \equiv 381=: T$ for all $i$. On the grid, we can write the predictor as $\eta_{il} = \sum_{r=1}^R f_r(\mathbf{x}_{rit_l},t_l)$. 
Each of the $R$ terms $f_r(\mathbf{x}_r,\mathbf{t})$,  containing evaluations $f_r(\mathbf{x}_{rit_l},t_l)$, $l=1, \dots, T_i, i=1, \dots, n$ in a vector, can then be represented through a tensor product
basis function expansion
\[
f_r(\mathbf{x}_r,\mathbf{t})\approx(\boldsymbol{\Phi}_{xr}\odot \boldsymbol{\Phi}_{tr})\boldsymbol{\vartheta}_r=\boldsymbol{\Phi}_{r}\boldsymbol{\vartheta}_{r},
\]
where $\mathbf{A}\odot\mathbf{B}=(\mathbf{A}\otimes\mathbf{1}_b\T)\cdot(\mathbf{1}_a\T\otimes\ \mathbf{B})$ denotes the row tensor product of the matrices $\mathbf{A}$ ($h\times a$) and $\mathbf{B}$ ($h\times b$), with $\mathbf{1}_d$ the $d$-vector of ones and $\cdot$ the  element-wise multiplication, and $\boldsymbol{\Phi}_{xr}$ and $\boldsymbol{\Phi}_{tr}$ contain the evaluations of the ($K_{xr}$ respectively $K_{tr}$) marginal basis functions for the covariate effects and over ${t}$, respectively. The effect shape is determined by $\boldsymbol{\vartheta}_r$ representing an unknown vector of coefficients. To provide sufficient flexibility of the model, the approximation uses a large set of basis functions, which is further regularised by a corresponding quadratic penalty term for the coefficients in the  (quasi-) log-likelihood
\[
\pen(\boldsymbol{\vartheta}_r|\lambda_{tr},\lambda_{xr}) = 
\boldsymbol{\vartheta_r}\T(\lambda_{xr}\mathbf{P}_{xr}\otimes\mathbf{I}_{K_{tr}}+\lambda_{tr}\mathbf{I}_{K_{xr}}\otimes\mathbf{P}_{tr})\boldsymbol{\vartheta_r}.
\]
Here, $\lambda_{xr}$ and $\lambda_{tr}$ are positive  smoothing parameters,  $\mathbf{P}_{xr}$ and $\mathbf{P}_{tr}$ are known and fixed positive (semi)definite marginal penality matrices corresponding to the basis matrices   $\boldsymbol{\Phi}_{xr}$ and $\boldsymbol{\Phi}_{tr}$, and $\mathbf{I}_{K_{xr}}$ and $\mathbf{I}_{K_{tr}}$ are identity matrices of dimensions $K_{xr}$ and $K_{tr}$, respectively
\citep[see][]{Scheipl:FAMM,FGAMM}. The unknown coefficients can then be estimated through a penalised (quasi-)maximum likelihood approach, maximising 
\[
\ell_p(\boldsymbol{\vartheta},\boldsymbol{\lambda},\boldsymbol{\nu}|\mathbf{y})=\ell(\boldsymbol{\vartheta},\boldsymbol{\nu}|\mathbf{y})-\frac{1}{2}\sum^R_{r=1}\pen(\boldsymbol{\vartheta}_r|\lambda_{tr},\lambda_{xr})
\]
where $\boldsymbol{\lambda}=(\lambda_{t1},\lambda_{x1}\ldots,\lambda_{tR},\lambda_{xR})$, 
$\boldsymbol{\vartheta}=(\boldsymbol{\vartheta}_1\T,\ldots,\boldsymbol{\vartheta}_R\T)\T$,  and $\ell(\boldsymbol{\vartheta},\boldsymbol{\nu}|\mathbf{y})$ is the (quasi-)log-likelihood implied by the respective response distribution \citep[see][]{FGAMM} optionally depending on additional nuisance (e.g.\ dispersion) parameters $\boldsymbol{\nu}$.

\subsubsection{Basis function representations for different  covariate effects}

The marginal basis matrices and corresponding penalty matrices are suitably chosen depending on the specified covariate effects. A full description is given in 
\cite{Scheipl:FAMM,FGAMM}; we here list some common choices also used in the model for the Covid-19 data for illustration and completeness, restricting to the case of equal grids for ease of presentation.
For covariate effects that are constant over $t$,  $\boldsymbol{\Phi}_{tr}=\mathbf{1}_{nT}$ is a vector of length $nT$ containing ones and $\boldsymbol{\mathbf{P}}_{tr}=\mathbf{0}$, while smooth time-varying effects are achieved when choosing $\boldsymbol{\Phi}_{tr}$ as a matrix of spline evaluations with $\boldsymbol{\mathbf{P}}_{tr}$ a corresponding penalty matrix (e.g.\ based on finite differences of B-spline coefficients).  Functional intercepts $\beta_0(t)$ are obtained through  $\boldsymbol{\Phi}_{xr}=\mathbf{1}_{nT}$ and $\mathbf{P}_{xr}=\mathbf{0}$. For effects $x\beta$ and $x\beta(t)$ that are linear in $x$, $\boldsymbol{\Phi}_{xr}$ changes to $\boldsymbol{\Phi}_{xr}=\mathbf{x}\otimes\mathbf{1}_T$ where $\mathbf{x}=(x_1,\ldots,x_n)\T$ and  $\mathbf{P}_{xr}=\mathbf{0}$. In case of a nonlinear effect specification for $x$, i.e.\ $f(x)$ and $f(x,t)$, $\boldsymbol{\Phi}_{xr}$ corresponds to a suitable marginal spline  basis matrix over $x$ and  $\mathbf{P}_{xr}$ is specified accordingly. 

For linear effects of a functional covariate $x(s)$, $s \in \mathcal{S}$, a tensor product spline representation for $\beta(s,t)$ is used  with marginal spline basis functions $\Phi_{k_s}, k_s=1,\ldots, K_{xr}$ over $\mathcal{S}$ and $\Phi_{k_t}, k_t=1,\ldots, K_{tr}$ over $\mathcal{T}$. 
This yields 
\[
\int_{\mathcal{S}}x_i(s)\beta(s,t_l)ds
\approx 
\int_{\mathcal{S}} x_i(s)\sum^{K_{xr}}_{k_s=1}\sum^{K_{tr}}_{k_t=1} \Phi_{k_s}(s)\Phi_{k_t}(t_l)\vartheta_{r,k_s,k_t} ds.
\]
Then  $\mathbf{\Phi}_{tr}=\left[\Phi_{k_t}(t_l)\right]_{\substack{l=1,\ldots,T\\k_t=1,\ldots,K_{tr}}} \otimes\mathbf{1}_n$
and 
$\mathbf{\Phi}_{xr} = 
\left[\int_{\mathcal{S}}x_i(s)\Phi_{k_s}(s)ds\right]_{\substack{i=1,\ldots,n\\  k_s=1,\ldots K_{xr}}}\otimes\mathbf{1}_T  
$
with marginal penalty matrices corresponding to the chosen marginal spline bases.
In practice, the integral is approximated using numerical integration.
A concurrent effect $f(x(t))$ or $f(x(t),t)$ of a functional covariate $x(s), s \in \mathcal{T},$ is constructed analogous to $f(x,t)$ above.

Finally, functional random intercepts for groups $c=1, \dots, M$, $c(i)$ being the group level of observation $i$, 
are associated with a marginal basis $\mathbf{\Phi}_{xr}=\left[\delta_{c(i)m}\right]_{\substack{i=1,\ldots,n\\m=1,\ldots,M}}\otimes\mathbf{1}_T$, with $\delta_{cm}$ the indicator for $c=m$ . 
The matrix $\mathbf{P}_{xr}$ then is a $M\times M$ precision matrix defining the dependence structure between levels of  $c$.





 
\subsection{Compositional predictor}\label{sec:coda}

We now introduce the new compositional predictor $\eta_i^{comp}(t)$. 
We first discuss existing methods for the case of finite compositions as covariates and scalar responses in Section  \ref{sec:sconcoda}, before introducing the proposed extensions to functional compositional covariates and/or (generalised) functional responses in 
\ref{sec:funonfuncoda}.

\subsubsection{Finite compositional covariates and scalar responses}
\label{sec:sconcoda}

Making use of the core principles of compositional data analysis \citep{10.5555/17272}, we formalise scalar-valued covariates which describe $D$ parts of a whole summing to a constant - such as the regional smoking status proportions - as compositions of $D$ parts living on the simplex
\[
\mathds{S}^D=\lbrace \mathbf{x}=(x_1,\ldots,x_D)^\top:x_d>0, d=1, \dots, D;\sum_{d=1}^Dx_d=\kappa\rbrace. 
\]
This space, equipped with the perturbation $\mathbf{x}\oplus \mathbf{y}=\cls(x_1y_1,\ldots x_Dy_D)$ and the powering $\alpha\odot\mathbf{x}=\cls(x_1^\alpha,\ldots,x_D^\alpha)$ operations, where $\mathbf{x,y}\in\mathds{S}^{D},\ \alpha\in \mathds{R}$ and \linebreak $\cls(\mathbf{x})=\left(\kappa x_1/\sum^D_{j=1}x_j,\ldots,\kappa x_D/\sum^D_{j=1}x_j\right)\T$ is the closure operator, as well as the inner product $\langle\mathbf{x,y}\rangle_A=(2D)^{-1}\sum_d\sum_j \log(x_d/x_j)\log(y_d/y_j)$,
is provided with a finite  $(D-1)$-dimensional Euclidean vector space structure  isometric to the $\mathds{R}^{D-1}$  \citep[cf.\ e.g.][]{PawlowskyGlahn2001}. Noting this isometric correspondence, a central idea in compositional data analysis is to map the provided relative information isometrically to $\mathds{R}^{D-1}$, perform  well-established statistical analysis methods there, and then potentially  back-transform the result onto  $\mathds{S}^D$ using inverse operations. 

Common operations include first the centred log ratio transform  
\[
\clr(\mathbf{x}) = \left[\log\frac{x_1}{m(\mathbf{x})},\ldots,\log\frac{x_{D}}{m(\mathbf{x})}\right],
\]
where $m(\mathbf{x})$ is the geometric mean of $\mathbf{x}$. The $\clr$  projects  the composition onto the $\clr$-plane $\mathcal{H}^D$, a $(D-1)$-dimensional subspace of $\mathds{R}^D$ whose components add to zero. 
By contrast, the second  transform $\ilr$ \citep{Egozcue2003} returns $(D-1)$ coordinates with respect to an orthonormal system on the $\clr$-plane  $\mathcal{H}^D$, which is  equivalent to the logit-function used in logistic regression for $D=2$. For $D>2$, 
infinitely many  orthonormal basis systems exist. As we will use the $\ilr$ only internally for estimation, the choice does not affect the interpretation and we use  
\textit{pivot coordinates} \citep{Fiserova2011}, 
for which the $(D-1)$-dimensional vector $\ilr(\mathbf{x})$ has components
\begin{equation}
\label{eq:ilrpivot}
{\ilr}_j(\mathbf{x})=\sqrt{\left(\frac{D-j}{D-j+1}\right)}\log\Bigg\{\frac{x_j}{\sqrt[D-j]{\prod_{k=j+1}^Dx_k}}\Bigg\},~j=1,\ldots,D-1.
\end{equation}

Making use of the isometric isomorphism between the Aitchison and the Euclidean geometry established through the $\clr$ and the $\ilr$ transformations, 
linear effects of compositional covariates can be modeled using 
$
\langle \mathbf{x,b} \rangle_A=\langle \clr(\mathbf{x}),\clr(\mathbf{b})\rangle=\langle \ilr(\mathbf{x}),\ilr(\mathbf{b})\rangle$  
\citep[e.g.][]{doi:10.1111/rssc.12283}. In particular, with 
$\mathbf{b}=\ilr^{-1}(\boldsymbol{\beta})$ the inverse of the regression coefficients in $\mathds{R}^{D-1}$, 
the compositional effect in the predictor is
\begin{equation}\label{eq:codapred}
 \langle\mathbf{b},\mathbf{x}\rangle_A = \sum^{D-1}_{j=1}\beta_j {\ilr}_j(\mathbf{x}).
\end{equation}
Recalling \cite{doi:10.1002/9781119976462.ch13}, $\mathbf{b}$ represents  the simplicial gradient of the predictor
with respect to the composition $\mathbf{x}$, and can be interpreted as the  direction of  perturbation of compositions on the $\mathds{S}^D$ which yields the largest effect on the outcome     \citep[cf.][]{doi:10.1111/rssc.12283}. That is, for $\mathbf{x}\oplus\mathbf{b}^\ast$ with  $\mathbf{b}^\ast=\mathbf{b}/\Vert\mathbf{b}\Vert_A$, we have 
\begin{equation} \label{direction}
\langle\mathbf{b},\mathbf{x}\oplus\mathbf{b}^\ast\rangle_A
=\langle\mathbf{b},\mathbf{x}\rangle_A+\Vert\mathbf{b}\Vert_A.
\end{equation}

Further, to quantify the effect of a change in the composition on the predictor, \cite{doi:10.1111/rssc.12283} suggested to perturb the composition into the direction of each part. For example, a change in the relative ratio  of the first compositional component of $\mathbf{x}$ by some $\alpha\neq 1$, while keeping the relative ratios for all other components constant, leads to a perturbation of  $\mathbf{x}$ by $\cls^{\prime}=\cls(\alpha,1,\ldots, 1)\T$ and a resulting change on the predictor by $\langle\mathbf{b},\cls^{\prime}\rangle_A 
=\log(\alpha)\clr_1(\mathbf{b})$, where $\clr_j$ is the the $j$-th component of the clr transformation, $j=1,\ldots,D$. In particular, for a log-link relation of the expected outcome and the predictor, as under the present model, the effect of a relative ratio change in the first component on the response scale simplifies to a change by the factor $\alpha^{\clr_1(\mathbf{b})}$.

\subsubsection{Extensions to functional compositional covariates and  functional responses} \label{sec:funonfuncoda}

The above formulation of \eqref{eq:codapred}  allows for a direct extension of the GFAMM to include compositional covariates by including the $(D-1)$ ilr-transformed coordinates as scalar covariates with linear effects,  such that $\boldsymbol{\Phi}_{xr}={\ilr}(\mathbf{X})\otimes\mathbf{1}_T$ and $\mathbf{P}_{xr}=\mathbf{0}$ for $\mathbf{X} = (\mathbf{x}_{id})_{i,d}$. Combination with suitable $\mathbf{P}_{tr}$ yields time-constant  $\langle\mathbf{b},\mathbf{x}\rangle_A$ or time-varying $\langle\mathbf{b}(t),\mathbf{x}\rangle_A$.

Treating density functions as infinite (functional) compositions, we extend the previous results to such covariates. 
The idea is to use an isometric isomorphism between the space of functional compositions and a subspace of the  $L^2$ space of functions via a functional $\clr$ transform, and to then 
treat the transformed functional composition as a functional covariate within the GFAMM framework using a suitably adapted basis function specification.

A suitable space in this context is the Bayes Hilbert space of densities  
\[
B^2(\mathcal{T})=\left\{f:\mathcal{T}\rightarrow (0,+\infty), \int_{\mathcal{T}} f(t) dt = 1,  \int_{\mathcal{T}}\left[\log(f(t)\right]^2dt<\infty\right\}
\]
\citep{doi:10.1111/anzs.12074}.
It generalises the Aitchison geometry from compositional data and provides a suitable geometric framework for the analysis of density functions. 
We here focus on some basic properties that are relevant in our setting and refer to \cite{doi:10.1111/anzs.12074} for a more formal definition and further mathematical details. Analogous to $\mathds{S}^D$,  
$B^2(\mathcal{T})$ has a vector space structure with perturbation and powering operations. For $f,h\in B^2(\mathcal{T})$, $t\in \mathcal{T}$ and  $\alpha\in\mathds{R}$, the perturbation $(\oplus)$ and powering  $(\odot)$ operations are defined by $(f\oplus h)(t)=f(t)h(t) / \int_{\mathcal{T}} f(t)h(t) dt$ and $(\alpha\odot f)(t)=f(t)^\alpha/ \int_{\mathcal{T}} f(t)^\alpha dt$, respectively.
Additionally, the inner product $\langle\cdot,\cdot\rangle_{B^2}$ on  $B^2(\mathcal{T})$  generalises the Aitchison inner product,
\[
\langle f,h\rangle_{B^2}=\frac{1}{2|\mathcal{T}|}\int_{\mathcal{T}}\int_{\mathcal{T}} \log\frac{f(t)}{f(s)}\log\frac{h(t)}{h(s)}dsdt
\]
where $f,h\in B^2(\mathcal{T})$ and $|\cdot|$ is the Lebesgue measure of the argument. In particular, noting that $\langle f,h\rangle_{B^2} = \langle \clr(f),\clr(h)\rangle_{L^2}$ where  $\clr(f)(t)=\log(f(t))-|{\mathcal{T}}|^{-1}\int_{\mathcal{T}}\log(f(s))ds$, 
the $B^2(\mathcal{T})$ can be shown to be a separable Hilbert space  and to be isometrically isomorph to the subspace $L^2_0(\mathcal{T})$ of functions in $L^2(\mathcal{T})$ integrating to zero with the usual $L^2$ metric. 
While this allows a transformation of densities to the $L^2(\mathcal{T})$, the additional integration-to-zero constraint of  $L^2_0(\mathcal{T})$ needs to be accounted for and, in general, prohibits a direct application of standard functional data analysis techniques to the transformed densities.

For the GFAMM, functional compositions $x(s), s\in \mathcal{S},$ are included into the regression with a linear effect in terms of the scalar product in $B^2$, using the equivalence \[
\langle x_i, b(.,t)\rangle_{B^2} = \langle \clr(x_i),\clr(b(.,t))\rangle_{L^2} = 
\int_\mathcal{S} u_i(s)\beta(s,t)ds
\]
with $u_i = \clr(x_i)$ and $\beta(.,t) = \clr(b(.,t))$ for each $t$. Note that $\beta$ is a surface with $\beta(.,t)\in L^2_0(\mathcal{T})$ 
fulfilling an integration-to-zero constraint for each $t$. Thus, we can estimate the effect similarly to a linear function-on-function regression term, with the modification of this additional constraint. We achieve this through the specification of a tensor product basis, which places an integration-to-zero constraint on the marginal basis for $\beta$ over $s$ 
(but not on the marginal basis over ${t}$) to get terms with integration-to-zero-for-each-t constraints \citep[see][Chapter  5.6]{Wood2006}.

In a post estimation step, the functional composition surface $b(s,t)$ with $b(.,t) \in B^2(\mathcal{T})$  for all $t$
can be computed through the  inverse $\clr$ transformation, 
$b(.,t) = \clr^{-1}(\beta(.,t))$ for each $t$. Similar to finite compositions, $b(.,t)$ can then be interpreted as the preferential direction in which to  perturb the functional composition to yield the largest increase in the outcome, from  $\langle x_i, b(.,t)\rangle_{B^2}$ to $\langle x_i, b(.,t)\rangle_{B^2}+\Vert b(.,t)\Vert_{B^2}$. 
\subsection{Model specification for Spanish Covid-19 incidence curves}

The regression model for our application includes scalar covariates  mostly  with time-varying effects (i.e.\ as linear concurrent functional terms). Based on expert opinion and the inspection of the roughly constant effect patterns, we considered the three lockdown indicators $x_{l,i},~l=1,2,3$, the rain indicator $x_{rain,i}$, the weekday indicators $x_{d,i},~d=1,\ldots,6$, the GPD $x_{gdp,i}$ and the transport system indicator  $x_{tra,i}$ to have time-constant effects. In constrast, recalling the hypothesised impact of overcrowded areas on the disease transmission, in particular during the initial stages of the pandemic, and the strong variation in the size  of the population over the different seasons for the coastal regions, we modelled the effect of the population density $x_{dens,i}$ and the coastline indicator $x_{sea,i}$ through linear concurrent  functional effects. To account for the observed time-varying patterns of the lagged temperature ($x_{temp,i})$, sun hours ($x_{sun,i}$), humidity ($x_{hum,i})$ and wind speed $(x_{wind,i})$ and the log transformed  non-zero precipitation ($x_{lprec,i}$) (summarised into  $x_{w,i}=(x_{temp,i},x_{sun,i}, x_{hum,i}, x_{wind,i},x_{lprec,i}))\T$, we considered a smooth, time-varying effect specification in an initial modelling strategy. However, as the estimated effect surfaces of all  covariates were roughly constant over time and monotone in the specific covariates, all weather covariates were specified to have time-constant smooth effects to reduce the complexity of the model. In addition to these terms,  we also considered a concurrent tensor product smoother for $x_{temp,i}$ and $x_{hum,i}$ to control for the interaction between these two  terms reported in the literature. To account for the observed spatial correlation among the provinces,  a spatially correlated functional random effect $\gamma_{i}(t)$ is included, using a MRF specification for the marginal basis $\mathbf{\Phi}_{xr}$. The structure for this MRF was derived from a Gabriel graph  \citep{matulaGabrielGraph}  to control for the strong economic and social interrelations of continental Spain and the Spanish islands and African enclaves. In addition, we included independent smooth functional random intercepts $\gamma_{0{com}_i}(t)$ for the $18$ community spatial    units to control for potential spatially nested effects and unobserved heterogeneity of the local Covid-19 measures on community level.  For the compositional covariates,  we included the effect of the smoker status  composition $\mathbf{x}_{smoke,i}=(x_{daily,i}, x_{occ,i},x_{ex,i}, x_{non,i})\T$ as a time-constant linear function-on-composition 
term (internally using the $\ilr$ transformation). The sex composition  $\mathbf{x}_{sex,i}=(x_{male,i}, x_{fem,i})\T$ effect was modelled with a time-varying linear function-on-composition term to account for the strong heterogeneity in proportions of males and females within the public health and the nursing sectors - with a clear majority of female workers - which yielded high numbers of infected females already at the beginning of the pandemic. Finally, for the age densities $x_{age,i}$ we considered a linear function-on-functional composition  term  (internally specified  through a tensor product interaction smooth of the $\clr$ transformed age curves). 
 
Combining these terms and writing $\boldsymbol{\kappa}_i=\{x_{it},t, \gamma_{i}(t),\gamma_{0,com_i}(t);  t \in \mathcal{T}\}$, the expected number of Covid-19 cases $\e\left[y_i(t)|\boldsymbol{\kappa}_i \right]$ 
for province $i$ is specified through the following regression equation 
\begin{eqnarray*}
\log\{\mathds{E}\left(y_{i}(t)|\boldsymbol{\kappa}_i\right)\} 
&=&~\log(N_{i}) + \beta_0(t) +  x_{rain,i}\beta_{rain} + x_{gdp,i}\beta_{gdp} + x_{tra,i}\beta_{tra}\\
&& + x_{sea,i}\beta_{sea}(t) + x_{dens,i}\beta_{dens}(t)
 + \sum^6_{d=1}x_{d,i}\beta_{d}
+ \sum^3_{l=1} x_{l,i}\beta_{l}\\  &&+\sum^4_{k=1}f_k(w_{k,i}(t-5))+x_{rain,i}(t-5)f_5(x_{lprec,i}(t-5))\\  
&& + f_6(x_{hum,i}(t-5),x_{temp,i}(t-5))
+ \gamma_{i}(t) +\gamma_{0,com_i}(t) \\
&&+ \langle \mathbf{x}_{smoke,i}, \boldsymbol{b}_{smoke}\rangle_A
+\langle \mathbf{x}_{sex,i}, \boldsymbol{b}_{sex}(t)\rangle_A
+ \langle x_{age,i}, b_{age}(.,t)\rangle_{B^2},
\\ 
\end{eqnarray*}
where $\log(N_i)$ is an offset for the population size $N_i$ in province $i$ and $(t-5)$ denotes a 5-day lag.

\section{Application to the Spanish Covid-19 data}\label{sec:results}

We now discuss the results of the proposed model for the Spanish Covid-19 data. All computations were performed in R version 4.1.1 \citep{Rcore},  using in particular the \texttt{compositions} \citep{RCoda} and  \texttt{refund} \citep{refund}  packages. The model was implemented using the \texttt{pffr()} function from the \texttt{refund}   package. Basis sizes before application of possible constraints are as follows. The marginal basis functions  of the smooth effects in the $t$ direction were specified through penalised B-splines \citep{10.1214/ss/1038425655} using $K_{tr}=30$ knots (yielding around 1 knot per 13 days), and $K_{tr}=28$ knots for the global functional intercept.  Covariate effect marginal bases were chosen smaller due to the smoothness of effects: $K_{xr}=10$ for the smooth effects of temperature and humidity,  $K_{xr}=9$ for sun hours and $K_{xr}=7$ for wind speed. Tensor product interactions were specified with $5 \times 5$ knots, which applied to the smooth interaction of temperature and humidity as well as to the function-on-function effect for clr(age). 

Under the above specification, the model 
explains  $96\%$ of the deviance. 
The estimated dispersion parameter under the quasi-Poisson specification is  15.1. Table \ref{tab:m1} shows the estimated covariate effects which are treated as time-constant, where the reported p-values are based on a Wald test  for the null hypothesis that each parametric term is zero \citep{Wood2006}.
All effects except for the indicators for the 2nd and 3rd global lockdown periods, and the 2nd ilr component are significant at a significance level of $\alpha=0.05$.  We found a negative effect of the first global lockdown period ({Lockdown 1}), indicating a reduction by around $12\%$  of the daily numbers of Covid-19 cases by the imposed measures (compared to the trend under no lockdown). 
\begin{table}
\caption{Estimated time-constant linear effects in the functional generalised additive model for the Spanish Covid-19 incidence\label{tab:m1}}
\centering
\fbox{%
\begin{tabular}{l*{5}{c}}
 & \em     $\beta$ &\em $\exp(\beta)$ &\text s.e. & $t$\text{-value} & Pr$(|T|>|t|)$\\
       \hline
Intercept & 3.44 & 31.12 & 1.30 & 2.65 & 0.008 \\ 
Lockdown 1 & -0.13 & 0.88 & 0.04 & -2.96 & 0.003 \\ 
Lockdown 2 & -0.02 & 0.98 & 0.14 & -0.16 & 0.875 \\ 
Lockdown 3 & 0.04 & 1.05 & 0.13 & 0.35 & 0.724 \\ 
ilr(smoke 1) & -5.13 & 0.01 & 1.73 & -2.98 & 0.002 \\ 
ilr(smoke 2) & -1.09 & 0.34 & 1.23 & -0.89 & 0.374 \\ 
ilr(smoke 3) & -5.79 & 0.00 & 1.96 & -2.95 & 0.003 \\ 
Monday & 0.34 & 1.41 & 0.01 & 33.81 & 0.000 \\ 
Tuesday & 0.41 & 1.51 & 0.01 & 41.00 & 0.000 \\ 
Wednesday & 0.37 & 1.45 & 0.01 & 36.11 & 0.000 \\ 
Thursday & 0.33 & 1.39 & 0.01 & 32.00 & 0.000 \\ 
Friday & 0.38 & 1.47 & 0.01 & 38.09 & 0.000 \\ 
Saturday & 0.14 & 1.15 & 0.01 & 13.09 & 0.000 \\ 
Transport & 0.45 & 1.56 & 0.01 & 29.53 & 0.000 \\ 
GDP & -0.37 & 0.69 & 0.02 & -15.29 & 0.000 \\ 
Rain & 0.05 & 1.05 & 0.01 & 4.47 & 0.000 \\ 
   \hline
  \end{tabular}}
\end{table}
The weekday effects show a clear positive impact on the numbers of Covid-19 notifications, which is similar for Monday through Friday and smaller for Saturday, compared to  Sunday. This heterogeneity over the weekdays may be due to  daily variation in the availability at local authority levels and of tests.
In line with the findings of \cite{paez2021}, the coefficient for \textit{transport} suggests an increase in Covid-19 cases by around $56\%$ if higher-order transit systems are available. The expected  incidence decreases with increasing GDP by around $31\%$ per 10.000 EUR, ceteris paribus. Lastly,  the rain indicator (representing days  with  non-zero levels of precipitation using a 5 day lag) shows a positive effect, leading to around  $5\%$ more Covid-19 cases after rainy days. 

\begin{figure}[h!]
\centering
\makebox{
       \includegraphics[scale=0.42]{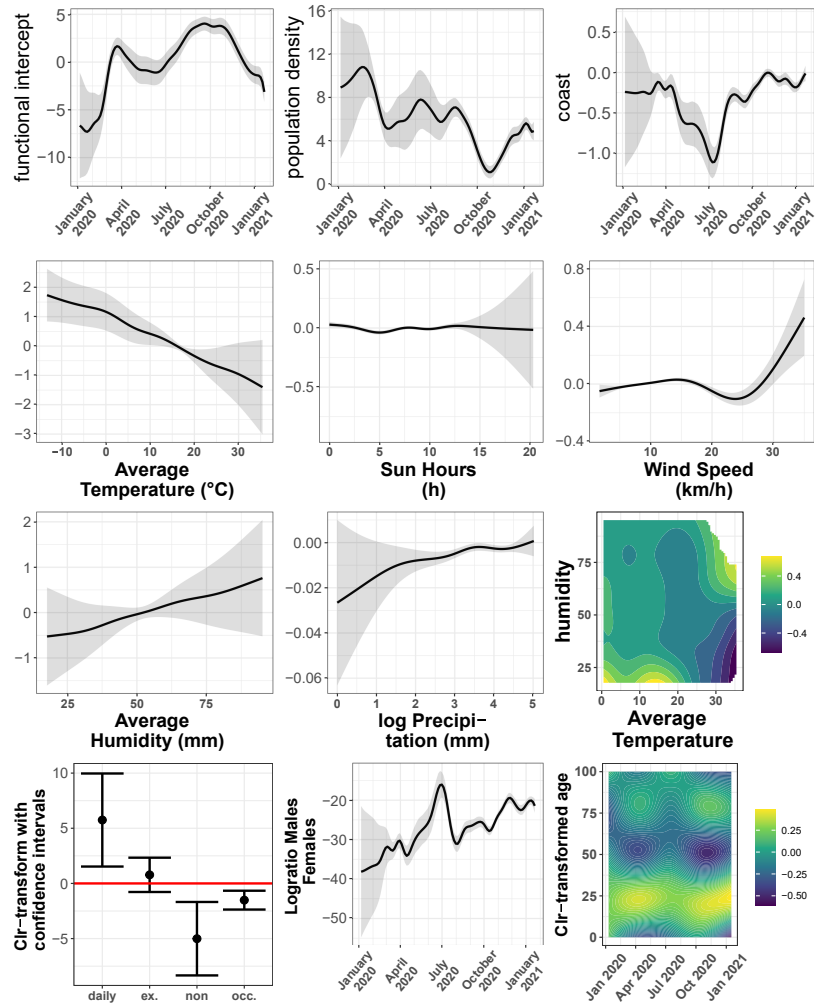}}
    \caption{Functional intercept and covariate effects: Smooth effects on expected number of daily Covid-19 cases. Upper row: functional intercept (left) and linear functional (time-varying) effects of the scalar covariates population density (central) and  coastline (right). Effects are given on the predictor level (on the log-mean) and pointwise 95\% confidence bands are shaded in gray. Central rows: nonlinear time-constant concurrent effects for different weather characteristics on the log-mean number of daily Covid-19 cases considering a 5 day lag. Upper panels: effects for lagged mean temperature, average sun hours, and maximum wind speed on the log-mean number of daily cases. Lower panels:  effects for the average lagged humidity,
    log transformed non-zero levels of the precipitation variable, and interaction effect surface for lagged mean temperature and lagged humidity (including main effect functions of temperature and humidity). Bottom row: 
    effects of the compositional covariates smoking status (left, clr-transformed), sex (middle, ilr-transformed) and age (right, clr-transformed) 
  on the log-mean number of Covid-19 cases.}
   \label{fig:mean}
\end{figure}

The first row of Figure \ref{fig:mean} shows the functional intercept and the linear functional (time-varying) effects of the scalar covariates.  The functional intercept (left)  has its highest peak during the second  Covid-19 wave, with maximum numbers of Covid-19 infections in mid-September. The effect for the population density at province level (central panel) reflects a clear positive impact on the expected number of infection notifications, with the strongest impact during the early stages of the first wave up to mid-March  2020. This finding is consistent with the  association of densely crowded areas with the  spread of the disease stated in the literature.
The  effect of \textit{coast} (right  panel) suggests smaller incidences for coastal compared to non-coastal provinces,  in particular  starting from mid-April 2020 onwards and reaching a minimum at around mid-July. This negative effect might be explained by an increased  public risk awareness and protective travelling behaviour caused by the aftermaths of the recent Covid-19 and lockdown experiences. Indeed, facing the massive impact of the disease on the Spanish population and health system during the first wave, overcrowded regions including the coast and metropolitan conurbations suffered a larger exodus of the population, and rural areas and the countryside became a favourite travelling destination. In addition, imposed national travelling restrictions and strict quarantine regulations for incoming and/or homecoming travellers yielded a strong reduction in numbers of international tourists and travellers. In a recent paper, \cite{SUN2021102062} reported a reduction of global scheduled flights for Spain by over $90\%$ for April to June 2020 compared to those month in 2019, which decreased to $65.7\%$ for July 2020. 
The observed negative effect of coast on the number of Covid-19 cases slowly vanishes towards the fall and winter of 2020, 
which could potentially be due to 
less protective individual travelling behaviour.      
  
\subsection{Concurrent functional effects of weather on Covid-19 cases}
The estimated  nonlinear time-constant concurrent effects of the lagged weather covariates and the interaction surface for the lagged mean temperature and lagged humidity  are depicted in the two central rows of Figure \ref{fig:mean}. For the lagged mean temperature (upper left panel), we found a negative effect of higher temperatures on the expected number of cases, which  is clearly supported by  \cite{WU2020139051} and  \cite{paez2021}.  The nonlinear effect for the lagged sun hours (upper central panel) only shows small positive and negative departures from zero, with confidence bands indicating high uncertainty especially for large values above 12 hours. The nonlinear effect for maximum wind speed (upper right panel) shows a small monotone increase in excepted daily notifications until around 15 km/h, with a decrease and increase for higher wind speeds becoming increasingly uncertain due to small numbers. The average humidity  and  the log transformed non-zero   precipitation values (lower left and central panels) show a roughly linearly increasing effect  on the expected incidence. Finally, the interaction surface of the lagged average temperature and the lagged humidity (lower right panel) suggests a  positive effect for low temperatures and low levels of humidity on the spread of the disease dynamics, contrasted with a negative impact of high temperatures and low humidity levels. The interaction indicates that smooth main effects of temperature and humidity should be interpreted with care, as they average over  parts of the interaction surface that depend on the data range. This might potentially explain the mixed results on the effects of  climate variables on the spread of the disease observed in different climatological regions, and is also seen in our sensitivity analysis below. 

\subsection{Compositional effect of smoking behaviour, sex and age on Covid-19 cases}

The effects of the compositional covariates on the expected number of daily notifications are shown in the final row of Figure \ref{fig:mean}. To interpret  the time constant effect of the individual smoking habits, we obtained the simplicial gradient via inverse $\ilr$ transformation of the corresponding 
coefficients on $\ilr$-level, see Table \ref{tab:m1}, indicating that the largest increase in the expected incidence is obtained by increasing the proportion of smokers (see Supplement, simplicial gradient roughly $(1,0,0,0)$). Applying the $\clr$ transformation to the simplicial gradient (see Section \ref{sec:sconcoda}) allows to evaluate the effect of a multiplicative change in the relative ratio of one component  while holding all other ratios constant. 
Depicted as sum-to-zero constrained effect estimates for the $\clr$ transformed composition  with corresponding $95\%$ confidence intervals, we found a clear positive effect of a larger fraction of daily (and less so ex-smokers) on the disease incidence, contrasted with a negative effect of a larger fraction of occasional and in particular non-smokers (see bottom left panel) which is in line with the results of \cite{Hopkinsonthoraxjnl} and \cite{Guelsen2020}. A relative ratio increase by $10\%$ for 
daily smokers yields a multiplicative increase in the expected  daily Covid-19 incidence by the factor $1.1^{5.747}=1.729$, i.e.\ by $ 73\%$. An analogous $10\%$ relative ratio increase for non-smokers ($1.1^{-5.009}$)  yields 
a 38\% decrease in the expected incidence. 

The estimated effect of the sex log-ratio (central panel) suggests a negative effect of an increase in the  male-to-female  ratio on the mean Covid-19 counts, particularly early in the pandemic. A possible explanation could be the described heterogeneity among the sexes in terms of employment in high-contact jobs such as in  the  retail and medical fields.

The right panel of the final row shows the effects of the clr-transformed age compositions on the disease dynamics, with estimated effect surface  constrained to fulfill an integration-to-zero constraint, $\beta(.,t)\in L^2_0(\mathcal{S})$, for each time point $t$. The effect surface shows clear variation over time and over the different ages. The  strongest positive effects on the number of Covid-19 cases appear for the younger and also for the very old parts of the population, with a clear mode for around 25 year olds. 
To interpret the effect of the age distribution on the incidence, we applied the inverse clr transformation to the estimated surface $\beta(.,t)\in L_0^2(\mathcal{S})$ for each $t$, to obtain for each time point the  direction $b(.,t)\in B^2(\mathcal{S})$  of change in the age composition leading to the largest increase in the mean incidence analogously to \eqref{direction}. Inspecting the time trend of $b(.,t)$  depicted in Figure~5 of the Supplement, all age curves show a clear mode for the younger ages and a second, but smaller, mode for around 80 year olds, with small variations in the exact density shape over time. This suggests that  provinces with high proportions of young people (and to a lesser extend old people) are more strongly affected by Covid-19 cases (see   Supplement for a discussion of the results). 

\subsection{Spatio-temporal effects}

A discussion of the results for both the  spatially correlated functional random intercepts per province and the spatially uncorrelated  community-specific functional random intercepts is given in the Supplement. Both effects  exhibit some variation in sign and effect size over the 52 provinces and 19 communities, respectively (see Figure 4 in the Supplement). 

\subsection{Model diagnostics and sensitivity analyses}

Comparing the fitted and the observed incidence curves (Figure 6 in the Supplement) shows only small deviations of the fitted from the observed daily values over the study period. The scaled Pearson residuals and the autocorrelation (Figure 7 in the Supplement) suggest overall good model fit, with  some amount of heterogeneity in variation and autocorrelation of the residuals, in particular some structure corresponding to the three Covid-19 waves, remaining.

We also performed sensitivity analyses to assess the effects of considering different lags for the weather covariates as well as a different  spatial neighbourhood specification. Overall, results were largely similar and general conclusions did not change (see the Supplement for a detailed discussion).  Note that while the interaction surface for temperature and humidity was stable under different model specifications and data sets, the main effects did change when considering continental Spain only, due to differences in the variable distribution over which the main effects average the interaction surface.

\section{Conclusion}\label{sec:final}

This paper has extended the generalised functional additive mixed model to the case when some  of the covariates in the predictor are finite or infinite (functional) compositions summing or integrating to a whole. We use an equivalence between the scalar product in the Bayes Hilbert space and a constrained linear (functional) term  for the clr transformed compositional covariate to embed the new model terms into the existing model framework.  For the transformed functional composition, the linear effect was modelled with a tensor product basis with a bivariate spline for the effect function, placing centering constraints on the marginal basis for the  covariate effect to account for the integration-to-zero constraint.
Although not considered here due to the increased model complexity given the sample size, the proposed model in principle also allows  to include  nonlinear  effects of the transformed (functional) covariates.  

The proposed model was applied to spatially correlated daily Covid-19 count data for Spain to quantify potential impacts of  population compositional, socio-economic, weather and regional covariates on the disease dynamics. The sampled data was collected from 5th January 2020 up to 19th January  2021, just before a large scale nationwide immunisation program was imposed in February 2021, which minimises  unknown effects on our results of the regionally varying and heterogeneous vaccination regimes. The available data has some limitations. First,  the reported numbers on a given day likely  represent  a mixture of counts over neighbouring (unobserved) true dates of symptom onset, given that  the reconstruction of symptom onset dates by a three day lag from the positive test results is only an  approximation. There is even more uncertainty regarding the true infection date, even if the average incubation period is 5 days. We may thus underestimate the weather effects, if the lagged weather variables only approximately measure weather on the date of infection. However, we did not detect large variations of the estimated results in our sensitivity analysis considering alternative lag specifications.
Second, the data does not provide separate infection counts for subgroups of the population according to sex, age and smoking habits. While we incorporate the effects of these variables on overall infections via compositional covariates measuring the composition of the population, we have to acknowledge the typical risks of ecological inference here. For instance, for the increasing effect of a larger share of smokers in the population on the Covid-19 incidence, we cannot distinguish whether this is due to a higher infection risk for smokers or due to a higher risk of Covid-19 positive smokers to infect others. 
Third, while the available data for 52 provinces and 381 days has better temporal and spatial resolution than other publicly available data sets, the data size still limits the complexity of the model in terms of the number of non-linear and/or time-varying effects. 
Taking these limitations into account, our model highlights a clear effect of the population composition according to sex, age and smoking habits, of weather variables (rain,  temperature, wind speed and  humidity), of GDP, population density, coast and public transit on the number of Covid-19 notifications, as well as spatial and temporal heterogeneity.

\clearpage
\begin{center}
\textbf{\large Supplemental Materials to Eckardt, Mateu and Greven:  Generalised functional additive mixed models with compositional covariates for areal  Covid-19 incidence curves}
\end{center}
\setcounter{equation}{0}
\setcounter{figure}{0}
\setcounter{table}{0}
\setcounter{page}{1}
\setcounter{section}{0}
\makeatletter

\newpage

\section{Variable description}

Table \ref{tab:variables_list} gives a description of the covariates used in the model for the Spanish Covid-19 incidence curves.

\begin{table}
\caption{\label{tab:variables_list}Description of the variables contained in the compiled data set. All variables are reported at province level, with the exception of the smoking composition, which is reported at (the coarser) community level,  corresponding to the third and second levels of the  Nomenclature des unit{\'e}s territoriales statistiques (NUTS) classification, respectively.}
\centering
\fbox{%
    \begin{tabular}{*{3}{l}}
\em Type &\em Variable &\em Description \\
       \hline
Scalars & gdp & Average individual gross domestic product per capita\\
& & in 10,000 EUR \\ 
& population & Number of inhabitants  \\
& population density & Population density  \\
& transport & Indicator variable for provinces with access to a\\
& &  public mass transport system \\
& coast & Indicator variable for coastal provinces\\[2mm]
& community & Identification number at NUTS2 level \\
& & (18 autonomous communities) \\
& province & Identification number at NUTS3 level \\
& &  (52 provinces)\\[2mm]


Compositions & sex & Proportion of males and females \\ 
& smoke & Proportion  of daily, ex-,  occasional and non-smokers  \\
& age & Age pyramids (0-100 yrs)  \\[2mm]

Time-varying & day & Weekday from monday to sunday (reference)  \\
& lockdown & Categorical variable indicating three global lockdown  \\
&& periods (reference no lockdown)\\
& humidity & Daily average humidity in percent  \\
& rain & Indicator variable derived from daily  precipitation \\
&& indicating precipitation larger than zero\\
& log(precipitation) & log transformation computed from  positive values\\
&& of daily precipitation \\
& sun &  Daily solar exposure in number of hours over the\\
& & irradiance threshold of 120 W/m$^2$  \\
& temperature & Daily average temperature in degrees centigrade  \\
& wind & Daily maximal wind speed in km/h \\
\end{tabular}}
\end{table}

\begin{figure}[h!]
\centering
\makebox{
    \includegraphics[scale=0.65]{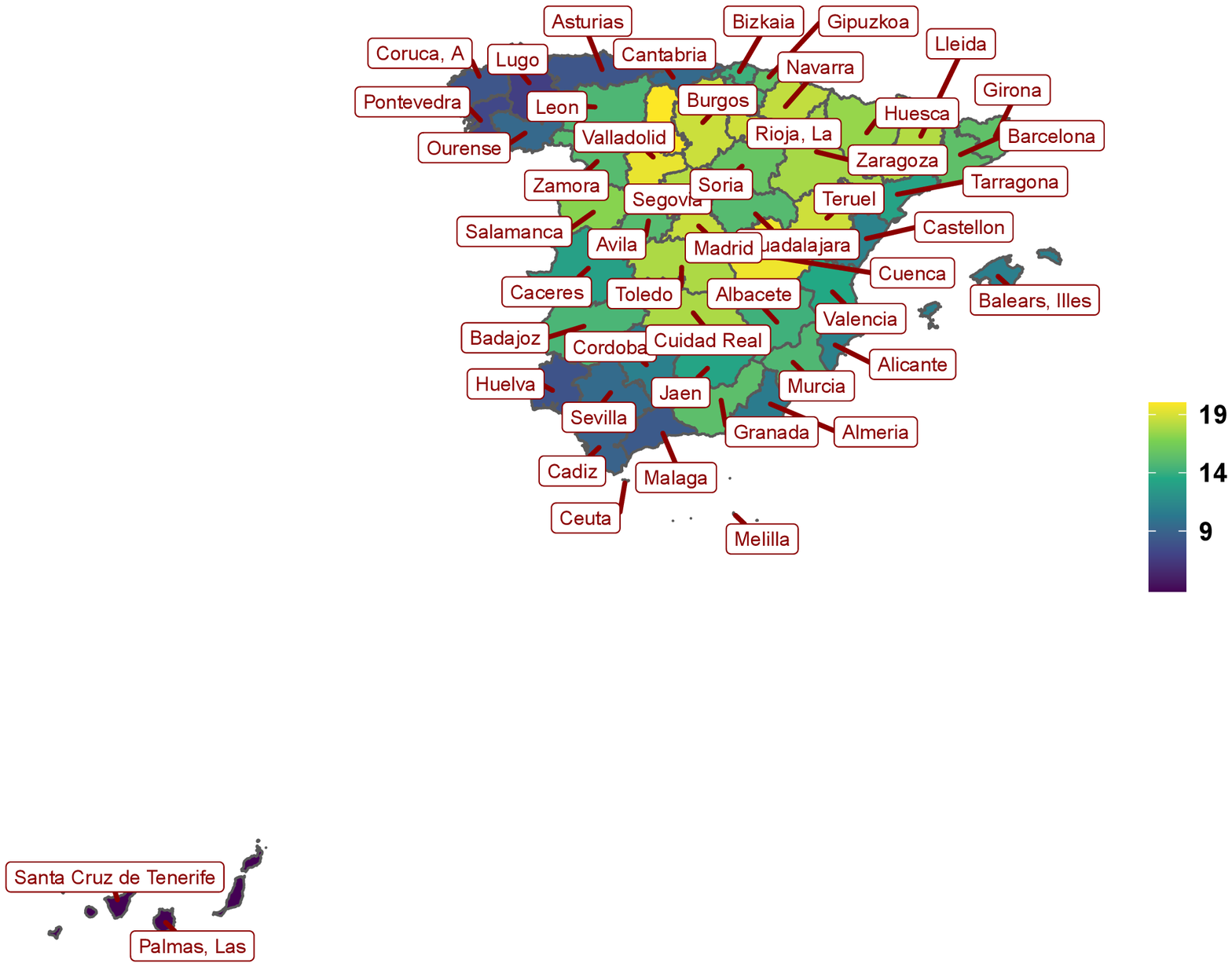}}
    \caption{Average Covid-19 incidence over 381 days at province (NUTS3) level.}
    \label{fig:allnuts2map}
\end{figure}

\begin{figure}[h!]
\centering
\makebox{
    \includegraphics[scale=0.65]{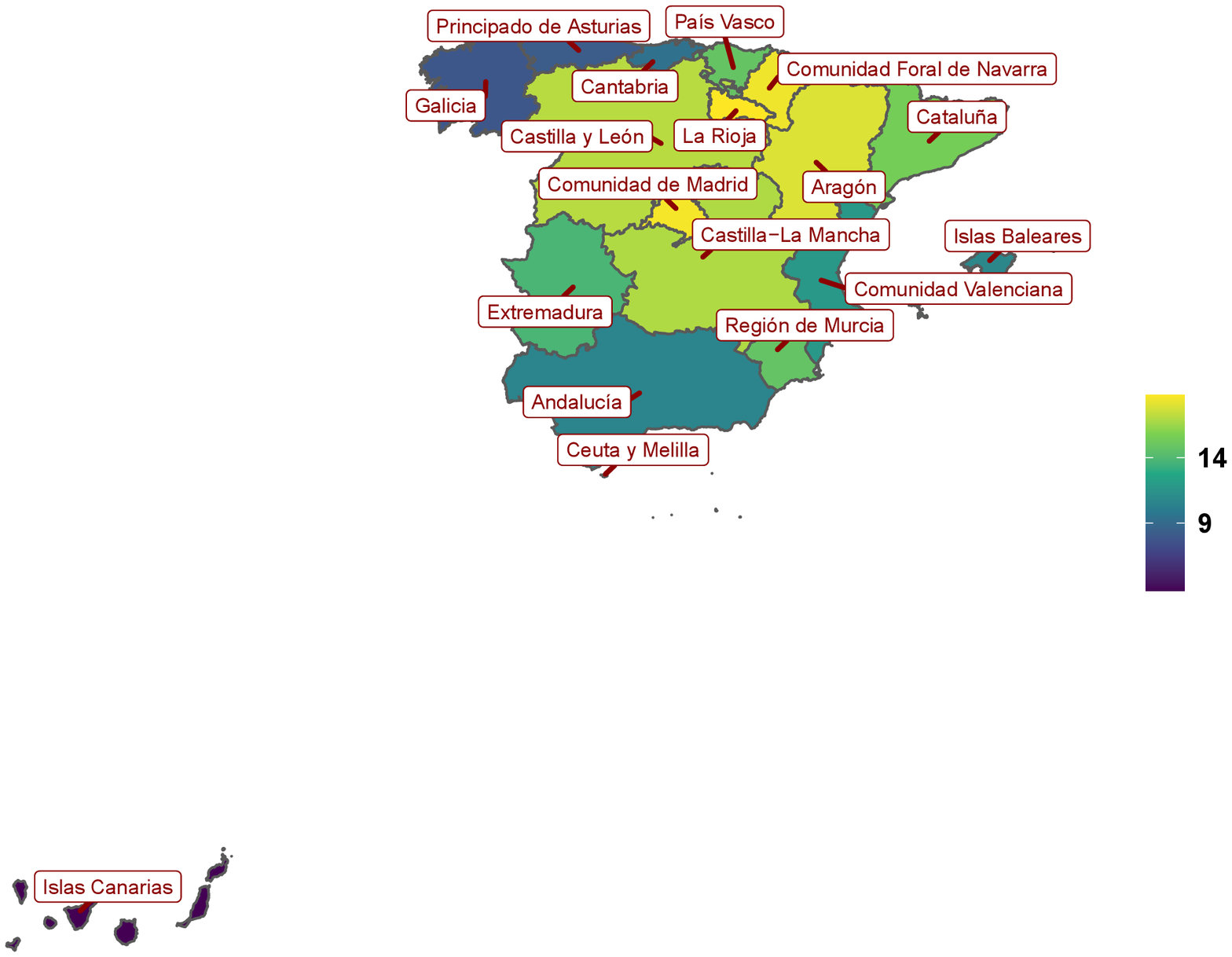}}
    \caption{Average Covid-19 incidence over 381 days at community (NUTS2) level.}
    \label{fig:nuts1map}
\end{figure}

\section{Additional Results}

\subsection{Spatio-temporal effects}

The spatially correlated functional random intercepts per province  and the spatially uncorrelated  community-specific functional random intercepts are shown in Figure \ref{fig:mrf_effect}. Both effects exhibit variation in sign and effect size over the 52 provinces and the 18 autonomous communities. 
The left panel of this plot for example show a clear positive deviation of the province of Lleida (red) from the overall intercept for the second wave in July 2020 contrasted with a negative deviation of Madrid (green).
The Markov Random Field basis used to specify this effect was derived from a Gabriel graph \citep{matulaGabrielGraph} specification using the \texttt{gabrielneigh} function of the \texttt{spdep} \citep{spdep_book, spdep_TEST} package in R \citep{Rcore}. This function generates a subgraph of a Delaunay triangulation constructed from the centroids of the individual areal units collected in a set of point locations $\mathbf{D}$. Within $\mathbf{D}$ two distinct locations $s$ and $s^{\prime}$ are said to be Gabriel neighbours if the closed disc with diameter equal to the distance between $s$ and $s^{\prime}$ does not contain any alternative point $z$ in $\mathbf{D}$. That is, $s$ and $s^{\prime}$ are connected whenever  $d(s,s^{\prime}) \leq \min_{z \in \mathbf{D}} ((d(s,z)^2 + d(s^{\prime},z)^2)^{1/2}).$ 
Different from the \emph{commonly shared border} specification, which does not allow  isolated regions, i.e. islands, and potentially yields a higher number of neighbours for the individual areal units, graph based approaches such as the Gabriel graph specification also work well for complex spatial structures including disconnected subregions or islands. Taking into account the strong economic and social interrelations of the continental provinces and the Spanish islands and African enclaves and the inter-regional interplay in the Spanish Covid-19 disease history, the Gabriel graph specification seems to preferable.  

\begin{figure}[h!]
\centering
\makebox{
    \includegraphics[scale=0.23]{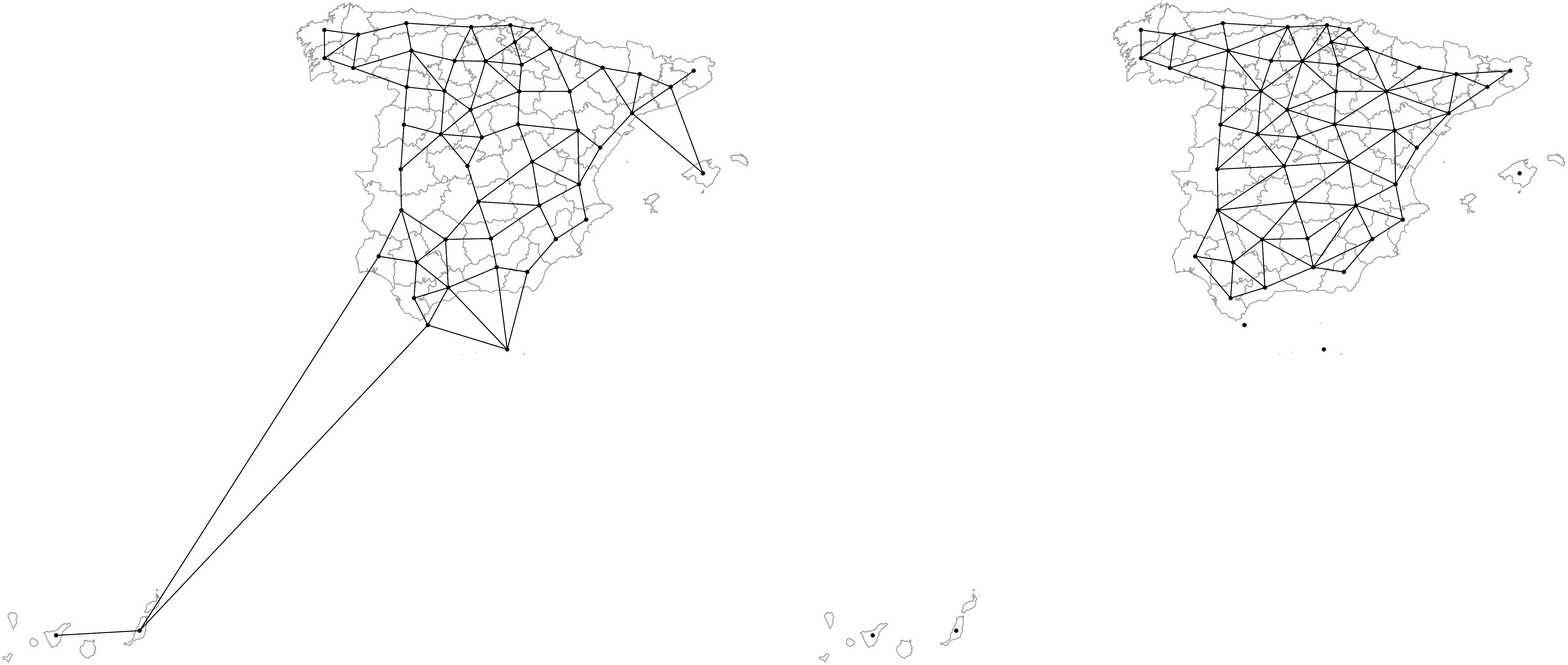}}
    \caption{Spatial neighbourhood structure of the spatially correlated functional random intercepts per province: Gabriel graph neighbourhood specification over all 52 provinces (left) and  commonly shared border based specifications defined over 47 continental provinces excluding the Balearic Islands, Las Palmas, Tenerife and the African enclaves Ceuta and Melilla.}
    \label{fig:graph}
\end{figure}

\begin{figure}[h!]
\centering
\makebox{
    \includegraphics[scale=0.5]{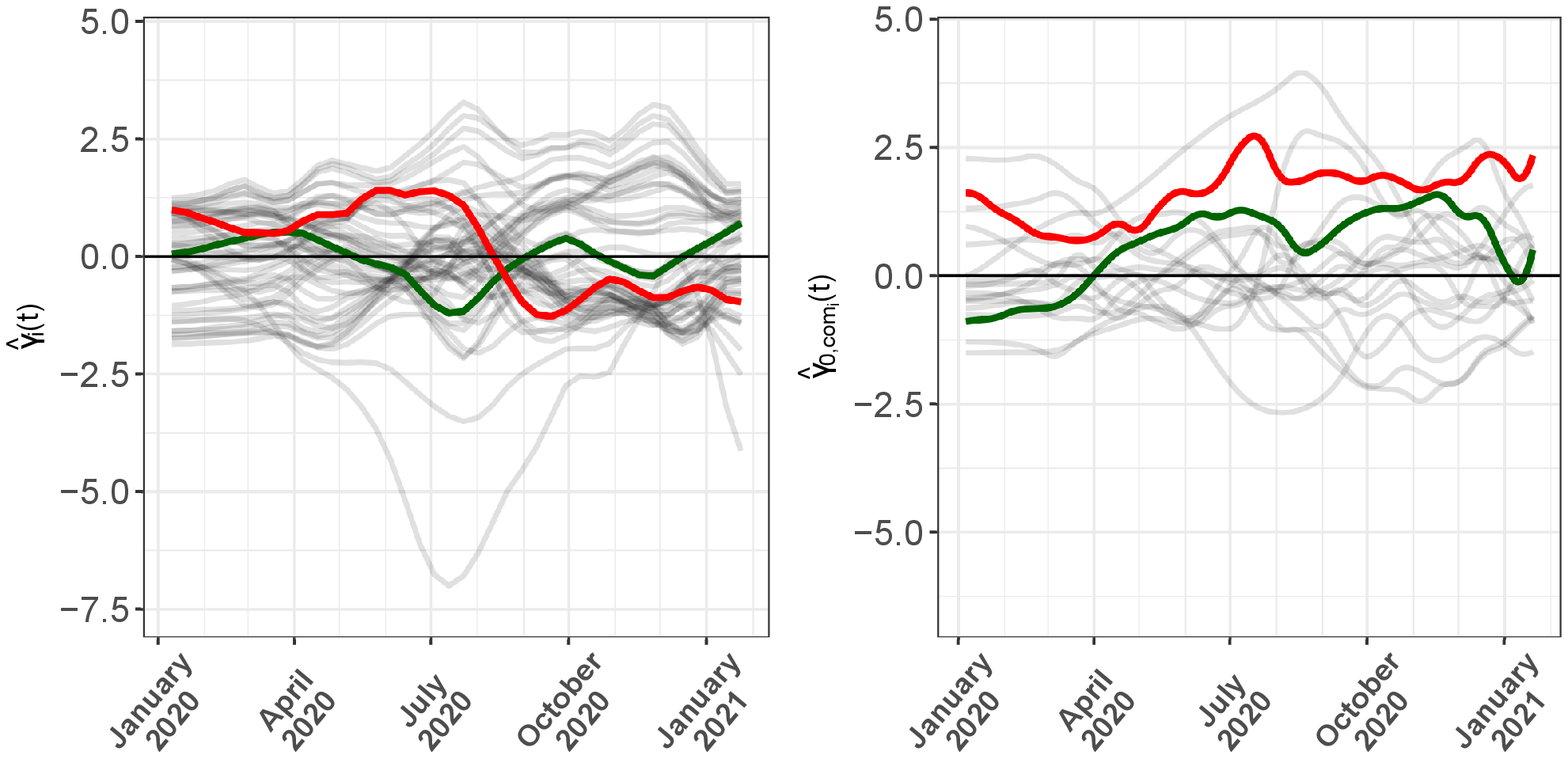}}
    \caption{Estimated functional random effects. Left:  spatially correlated functional random intercepts per region (Markov Random Field specification), with  estimated curves for the provinces of Madrid (green) and Lleida (red) highlighted. Right: spatially uncorrelated functional random intercepts per  community, with  estimated curves for the communities of Madrid (green) and Catalonia (red) highlighted.}
    \label{fig:mrf_effect}
\end{figure}

The constructed neighbourhood structures under the Gabriel and also the commonly shared border based definitions are visualised in Figure \ref{fig:graph}. Under the Gabriel graph specification, two provinces (Girona,  Tenerife) are  connected to one alternative province while, at  maximum, three  provinces (Avila, Burgos, Malaga) are connected to six alternatives provinces.   

The estimated spatially uncorrelated  functional random intercepts for the larger areal units \emph{communities} are depicted in the right panel of this plot. This effect  controls for unobserved spatial heterogeneity at the community level, where different local  measures were imposed after the global lockdown (captured through the three global lockdown indicators in our model) ended. For example, both communities Catalonia (red) and Madrid (green), containing the provinces of Lleida and Madrid, respectively, show clear positive deviations from zero for the second and third Covid-19 waves. 
The largest variation among the community-specific functional intercepts is shown during the end of the global lockdown period and early stages of the second Covid-19 wave from June and July 2020 onwards. These differences may  be due to differences in the regional measures which were imposed after the global lockdown at only the  local level.   

\subsection{Effects of the smoking composition}

The simplicial gradient $\mathbf{b}$ derived from the inverse $\ilr$ operation of the corresponding coefficients and its $\clr$ transformation are reported in Table \ref{tab:smoking_effects}. 
\begin{table}
\caption{Simplicial gradient and its $\clr$ transformation derived from the estimated $\ilr$ regression coefficients for the smoking composition \label{tab:smoking_effects}}
\centering
\fbox{%
\begin{tabular}{l*{4}{c}}
 & \text{daily smoker}& \text{occasional smoker}&\text{ex-smoker} & \text{non-smoker}\\
       \hline
$\mathbf{b}$ & 0.9924 & 0.0007 & 0.0069 & 0.0000 \\
$\clr(\mathbf{b})$ & 5.747 &  -1.515 &   0.778 &  -5.010 \\
\end{tabular}}
\end{table}

We estimated an  increase of the relative ratio for daily smokers by $10\%$ to increase the expected daily incidence by the factor $1.1^{5.747} = 1.729$ or about 73\%. To see what an increase in the  relative ratio for daily smokers by $10\%$ translates to, we can look at regional smoking compositions and at a perturbation of the original compositional vectors.  For example, applied to the regional smoking composition of Castell{\'o}n  (a province located at the Mediterranean coast in the eastern part of Spain), an increase in the  relative ratio for daily smokers by $10\%$ would change the compositional vector $\mathbf{x}_{Cast}=(0.248,0.021,0.205,0.527)\T$ (daily, occasional, ex-, non-smokers) to  $\tilde{\mathbf{x}}_{Cast}=\mathbf{x}_{Cast}\oplus \mathcal{C}(1.1,1,1,1)\T
=\mathcal{C}(1.1 \cdot {x}_{Cast,1}, {x}_{Cast,2},{x}_{Cast,3}, {x}_{Cast,4})\T=(0.266, 0.020,  0.200, 0.514)\T$.

\subsection{Effects of the age composition}

To interpret the estimated coefficient surface,  the $clr$ transformations of the simplical gradient $\mathbf{b}(.,t)$ of the age compositions for selected ages and dates are reported in Table~\ref{tab:clr:age}. Investigating the effect of a multiplicative change in the relative ratio of one functional component (age) at a particular date $t$, suggests that a 10\% relative ratio increase of 5 year olds while holding all other ratios constant yields a decrease in Covid-19 cases on March, 1st 2020  by the factor $1.1^{ -0.13} = 0.988$, i.e.\ by 1.2 $\%$. On the same date, we found an increase of 3$\%$ in case of a corresponding relative ratio increase of 25 year olds ($1.1^{0.34}=1.03$).  

\begin{figure}[h!]
    \centering
    \makebox{
    \includegraphics[scale=0.54]{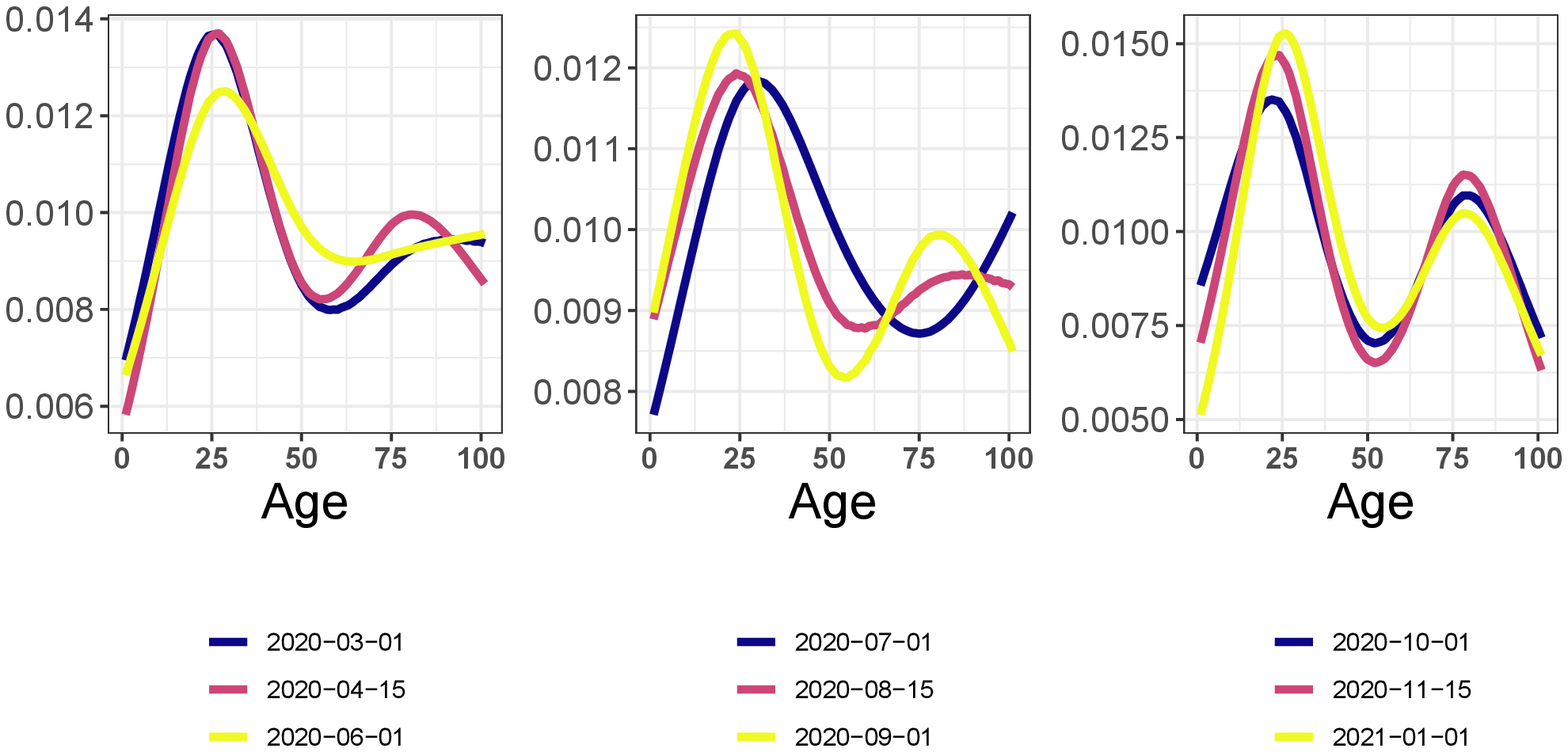}}
    \caption{Inverse clr transformed age effects on the Bayes Hilbert Space for ages 0 to 100 for selected dates.}
    \label{fig:m1_clr_age_curved_males}
\end{figure}

\begin{table}
\caption{\label{tab:clr:age} Clr-transformations of functional simplicial gradient for selected dates and ages}
\centering
\fbox{%
    \begin{tabular}{rrrrrrrr}
\em Age &\em 01.03.20 & \em 15.04.20 & \em 01.06.20 & \em 01.07.20 & \em 15.08.20  &\em 01.09.20 & \em 01.10.20 \\
       \hline
5 years & -0.13 & -0.26 & -0.22 & -0.13 & -0.01 & 0.02 & 0.03 \\ 
10 years & 0.05 & -0.03 & -0.05 & -0.03 & 0.07 & 0.12 & 0.17  \\ 
15 years & 0.21 & 0.16 & 0.08 & 0.06 & 0.14 & 0.19 & 0.28 \\ 
25 years & 0.34 & 0.34 & 0.24 & 0.18 & 0.19 & 0.23 & 0.31 \\ 
35 years & 0.22 & 0.23 & 0.20 & 0.17 & 0.11 & 0.10 & 0.08  \\ 
40 years & 0.10 & 0.10 & 0.14 & 0.14 & 0.04 & -0.01 & -0.08  \\ 
50 years & -0.15 & -0.14 & -0.02 & 0.02 & -0.09 & -0.17 & -0.32\\ 
60 years & -0.19 & -0.15 & -0.08 & -0.07 & -0.11 & -0.15 &-0.21\\
70 years & -0.12 & -0.04 & -0.08 & -0.12 & -0.08 & -0.04 & 0.03\\ 
80 years & -0.05 & 0.02 & -0.06 & -0.11 & -0.05 & 0.01 & 0.12 \\ 
85 years & -0.04 & 0.01 & -0.05 & -0.09 & -0.04 & -0.00 & 0.07\\ 
\end{tabular}}
\end{table}
In addition, we computed the simplicial gradient $\mathbf{b}(.,t)$ through the inverse $\clr$ operation of  the effect surface. The inverse $\clr$ transformed age effects for selected dates are depicted in Figure  \ref{fig:m1_clr_age_curved_males}. All panels support the above findings and highlight the relevance of the younger (and to a smaller extend also of the older) age groups on the spread of the disease.   
In particular, we found a clear increase of expected Covid-19 cases  for regions with high proportions of younger ages (mode around 25)  compared to only smaller increases in disease notifications for regions with larger proportions of above fifty year olds (mode around 75). The shape of the effect varies slightly over the different waves.  

\subsection{Fitted and observed Covid-19 curves}

\begin{sidewaysfigure}
\centering
    \makebox{
    \includegraphics[scale=0.53]{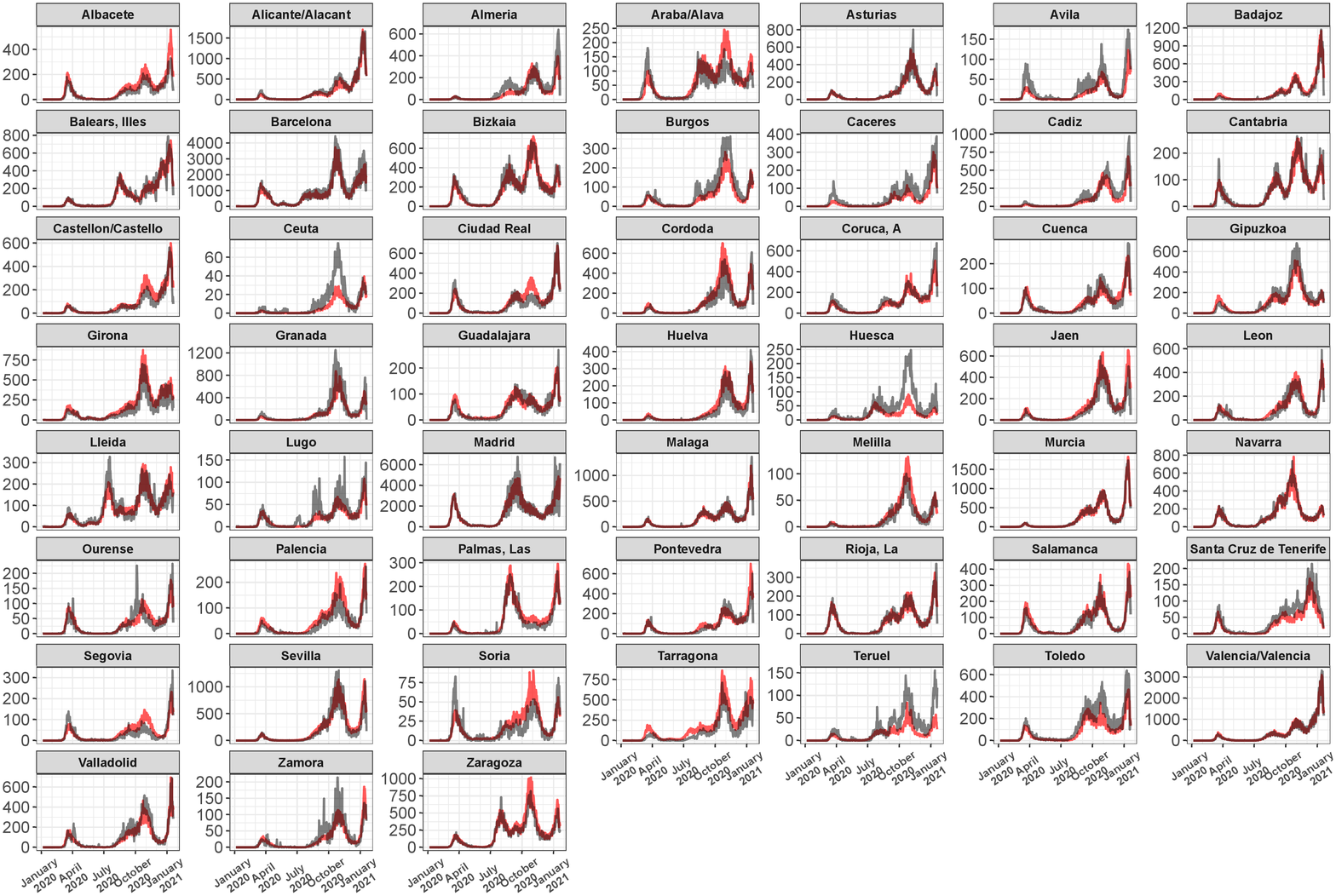}}
    \caption{Fitted (red) and observed (black) incidence curves all 52 provinces over the complete period under study.}
    \label{fig:m1_fit}
\end{sidewaysfigure}

The fitted against the observed response curves for all 52 Spanish provinces are plotted in Figure \ref{fig:m1_fit}. While most plots plots suggest small deviations of the fitted curves  (red) from the reported values (black), some regions show larger deviations between both curves including e.g. the provinces of \'{A}vila, Huesca and Lugo. 
For all three provinces, the model underestimates the reported numbers which might, in part, be explained by the neighbourhood structure. For example, located in northeastern Spain, the province of Huesca has a higher incidence rate in October compared to its   neighbouring provinces of  Lleida, Navarra, Tarragona and Zaragoza. Likewise, \'{A}vila located in central-western Spain has a higher incidence rate in March and August to November 2020  compared to most of its neighbouring provinces of Caceres,  Madrid, Salamanca, Segovia, Toledo and Villadolid.
The fit to the data could potentially be improved further by additionally including a province-specific spatially uncorrelated functional random intercept; however given the model complexity relative to the data size we refrain from this additional model extension.

\begin{figure}[h!]
  \centering
  \makebox{
    \includegraphics[scale=0.7649]{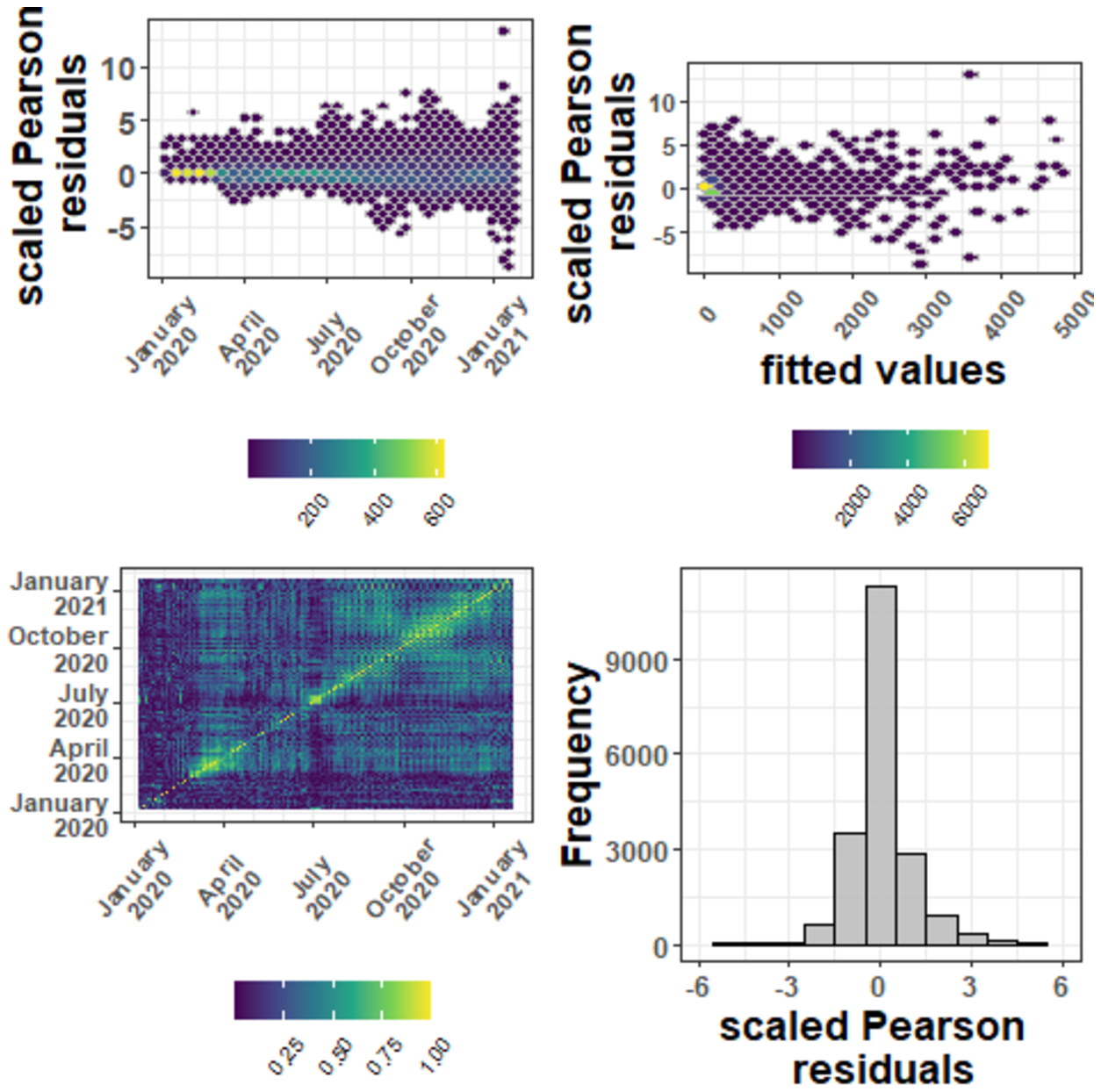}}
    \caption{Selected residual plots computed for the generalised functional additive model. Heatmap of binned points for scaled Pearson residuals against the fitted values (upper left panel), heatmap of binned points for scaled Pearson residuals over time (upper right panel),  autocorrelation of scaled Pearson residuals over time (bottom left), and histogram of scaled Pearson residuals (bottom right).}
    \label{fig:resid}
\end{figure}

Figure \ref{fig:resid} shows different  residual plots computed for the extended generalised functional additive mixed model. The computed scaled Pearson residuals of the upper panels and the corresponding histogram shown in the lower right panel of this plot suggest that some amount of heterogeneity in variation of the residuals remains. While a large number of the residuals takes values close to zero over the observed period (upper left panel), the pattern of the scaled Pearson  residuals against the fitted values shows some remaining heterogeneity over the functional domain with peaks corresponding to the three Covid-19 waves. Investigating the variance of the residuals, which should be around 1, we found that $20.99\%$ and $5\%$ of the scaled Pearson residuals have absolute value greater than 1 and 2, respectively, indicating no unusually high number of large values.   

The autocorrelation of the scaled Pearson residuals (lower left panel) also reflects some structure, again corresponding to the three Covid-19 waves. While the remaining structure could likely be reduced by additionally incorporating a spatially uncorrelated functional random intercept per province, by increasing the number of basis functions over time and/or including province-specific weekday effects, we refrained from these model extensions due to the increased model complexity relative to the sample size.

\section{Sensitivity analyses}

\subsection{Effect of the neighbourhood specification}

\begin{figure}[h!]
    \centering
    \makebox{
    \includegraphics[scale=0.52]{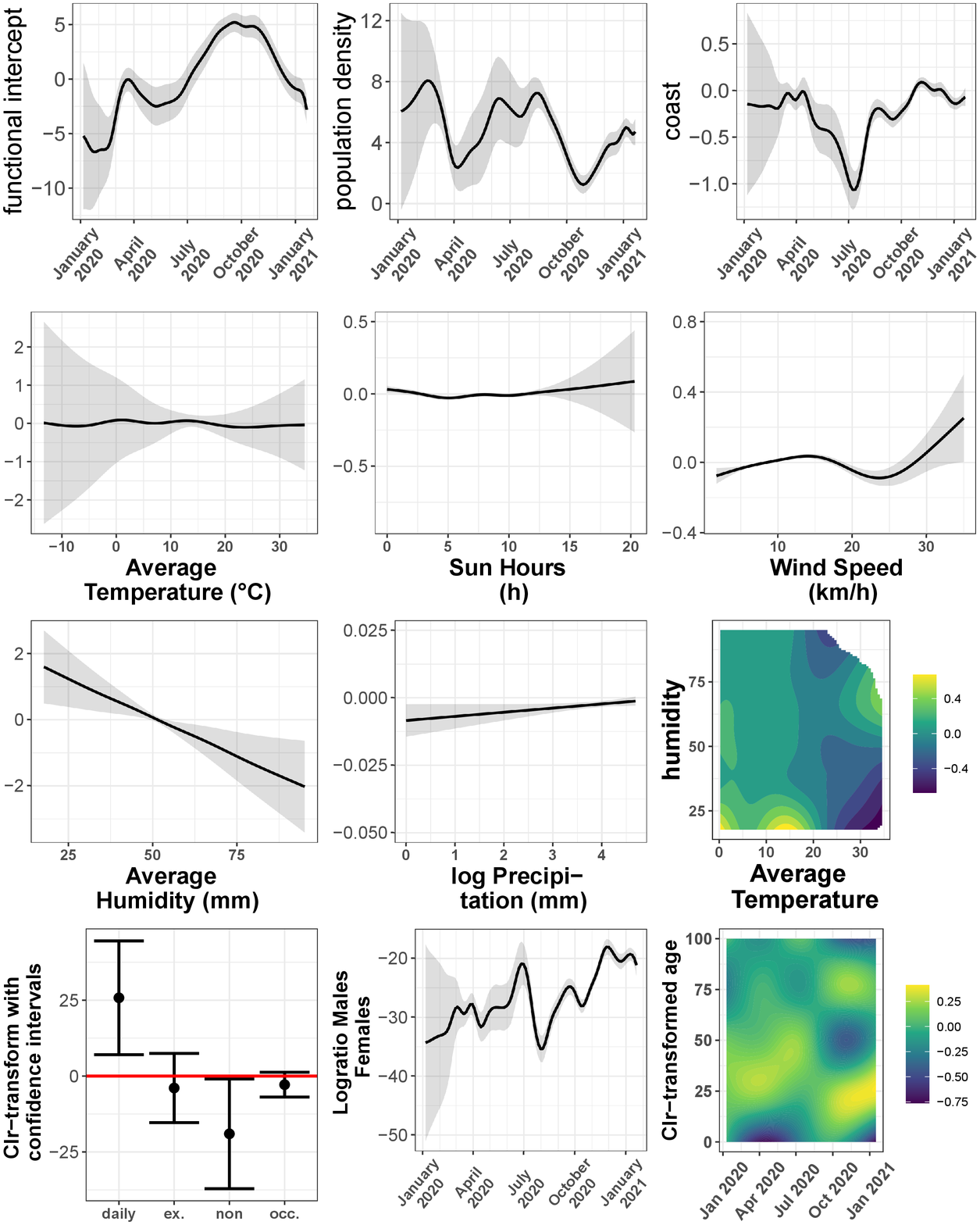}}
    \caption{Functional intercept and effects under the commonly shared border based neighbourhood definition.}
    \label{fig:m1_effect_nonline_queen}
\end{figure}

To investigate  the effect of the MRF specification in the extended GFAMM, we replaced the Gabriel graph neighbourhood specification of the spatially correlated functional random intercepts at province level through a commonly shared border neighbourhood approach. Different from the Gabriel graph approach used in the model discussed in the main document, this neighbourhood specification does not allow for isolated regions such as the African enclaves or the Spanish islands and we thus had to decrease the data set accordingly, leaving 47 provinces and 15 communities. The resulting model yielded an explained deviance of $96.2\%$ and an estimated scale parameter of $16.025$. In general, most effects of Table \ref{tab:m1:queen} and Figure \ref{fig:m1_effect_nonline_queen} have the same sign and are similar in size and shape. However, the effect of lagged humidity, lagged average  temperature and the smoking composition is somewhat different under the common shared borders approach. At the same time, the interaction surface  for the lagged average temperature and lagged humidity shows a similar effect pattern compared to the Gabriel graph based model. 
\begin{table}
\caption{Constant coefficients of the functional generalised additive model under the commonly shared border neighbourhood definition\label{tab:m1:queen}}
\centering
\fbox{%
\begin{tabular}{l*{5}{c}}
 & \em     $\beta$ &\em $\exp(\beta)$ &\em s.e. &\em $t$-value & Pr$(>|t|)$\\
       \hline
Intercept & 12.25 & 209133.29 & 5.04 & 2.43 & 0.01 \\ 
Lockdown 1 & -0.13 & 0.88 & 0.04 & -2.99 & 0.00 \\ 
Lockdown 2 & -0.06 & 0.94 & 0.15 & -0.41 & 0.68 \\ 
Lockdown 3 & -0.00 & 1.00 & 0.13 & -0.02 & 0.99 \\ 
ilr(smoke 1) & -20.23 & 0.00 & 8.21 & -2.46 & 0.01 \\ 
ilr(smoke 2) & -12.58 & 0.00 & 5.81 & -2.16 & 0.03 \\ 
ilr(smoke 3) & -21.97 & 0.00 & 10.65 & -2.06 & 0.04 \\ 
Monday & 0.35 & 1.42 & 0.01 & 32.71 & 0.00 \\ 
Tuesday & 0.42 & 1.52 & 0.01 & 39.70 & 0.00 \\ 
Wednesday & 0.38 & 1.46 & 0.01 & 35.06 & 0.00 \\ 
Thursday & 0.33 & 1.39 & 0.01 & 30.98 & 0.00 \\ 
Friday & 0.39 & 1.48 & 0.01 & 36.82 & 0.00 \\ 
Saturday & 0.14 & 1.15 & 0.01 & 12.56 & 0.00 \\ 
Transport & 0.66 & 1.92 & 0.02 & 42.09 & 0.00 \\ 
GDP & -0.70 & 0.49 & 0.03 & -25.56 & 0.00 \\ 
Rain & 0.03 & 1.03 & 0.01 & 3.73 & 0.00 \\ 
   \hline
\end{tabular}}
\end{table}
The observed difference might be explained through the  differences in the variable distributions of the distinct populations under study. Comparing the global averages computed over the complete study period and  all 52  provinces of Table \ref{tab:isolated:VS:continental} with the corresponding global mean values for the  continental and the non-continental provinces, we found that the Spanish islands and African enclaves show higher values for  temperature, wind speed, sun hours, humidity and also males, contrasted with lower average numbers in rain, GDP and the population density and, in particular, the average number of  daily Covid-19 cases. Investigating the smoking proportions of the complete data and the subset of continental and non-continental provinces, we found slight differences between the Spanish islands and African enclaves  provinces compared to continental Spain.  

\begin{table}
\caption{Global means and smoker proportions calculated over the complete study period and all 52 provinces, the subset of 47 continental and the subset 5 non-continental provinces  \label{tab:isolated:VS:continental}}
\centering
\fbox{%
\begin{tabular}{lrrr}
 & \em  all provinces    &\em continental provinces &\em isolated provinces\\
       \hline
Cases & 122.15 & 130.24 & 46.08 \\ 
Temperature & 15.77 & 15.35 & 19.71 \\ 
Rain & 0.31 & 0.32 & 0.18 \\ 
Average wind speed & 3.13 & 3.03 & 4.04 \\ 
Maximum wind speed & 9.73 & 9.71 & 9.93 \\ 
Sun hours & 7.32 & 7.27 & 7.81 \\ 
Humidity & 52.60 & 51.44 & 63.55 \\ 
log Precipitation & 0.41 & 0.43 & 0.22 \\ 
Population density & 0.01 & 0.01 & 0.00 \\ 
GDP & 2.25 & 2.27 & 2.06 \\ 
Males & 0.49 & 0.49 & 0.50 \\ 
Age & 43.99 & 44.11 & 42.53 \\
Daily smokers  & 0.22 & 0.22 & 0.23  \\
Occasional smokers & 0.02 & 0.02 &  0.03 \\
Ex-smokers & 0.25 & 0.25 &  0.22\\
Non-smokers & 0.51 &  0.50 &  0.53 \\
   \hline
\end{tabular}}
\end{table}

\begin{figure}[h!]
\centering
\makebox{
    \includegraphics[scale=0.4]{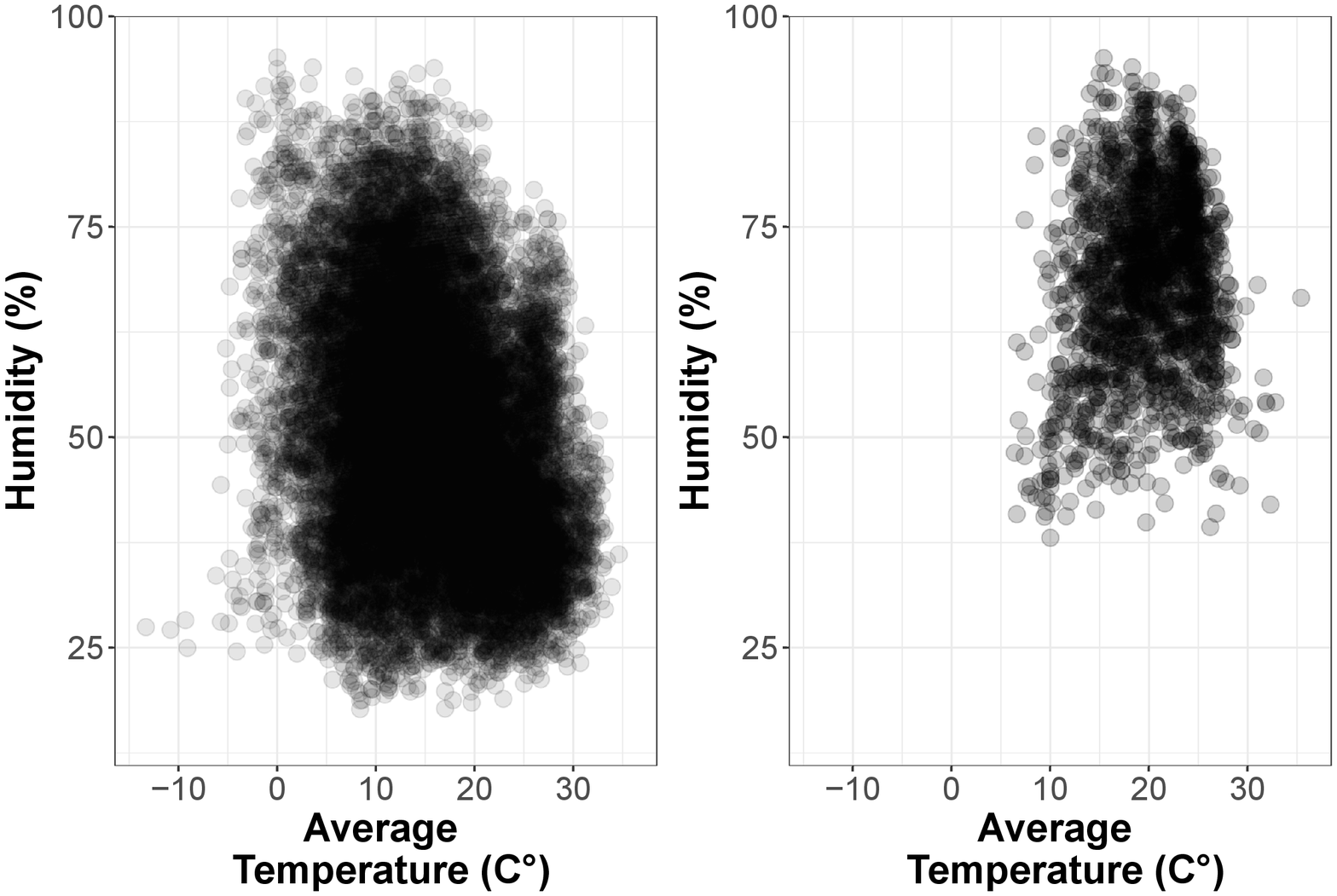}}
    \caption{Scatterplots of humidity and average temperature for 47 continental (left) and 5 non-continental Spanish provinces. 
    \label{fig:huteint}}
\end{figure}

To investigate potential  variations of the lagged average temperature and lagged humidity pattern between the continental and non-continental Spanish province, we plotted the joint distribution of both variables in Figure  \ref{fig:huteint}.
As suggested by Table \ref{tab:isolated:VS:continental}, both the recorded  temperature and humidity values are lower for the continental (left panel) Spanish provinces, ranging from $18\%$ to $95\%$ (humidity) and $-13.3^{\circ}$ to $34.6^{\circ}$ centigrade, compared to the non-continental provinces, ranging from $38\%$ to $95\%$ (humidity) and $6.4^{\circ}$ to $35.4^{\circ}$ centigrade, and the non-continental provinces in particular have high probability mass for both high temperature and humidity.

This difference in the bivariate distribution means that the main effects average over different parts of the interaction surface for humidity and temperature, leading to no main  effect  (continental provinces) vs. a negative effect (all provinces) for temperature and a negative main effect (continental provinces) vs. no main effect (all provinces) for humidity. This illustrates that smooth main effects are difficult to interpret in the presence of a smooth interaction.

Note that the differences in results are mainly due to the reduced data set and not to the neighbourhood specificiation, as we will see below when discussing results from a Gabriel graph based model for the reduced data set.

\begin{figure}[h!]
\centering
\makebox{
    \includegraphics[scale=0.5]{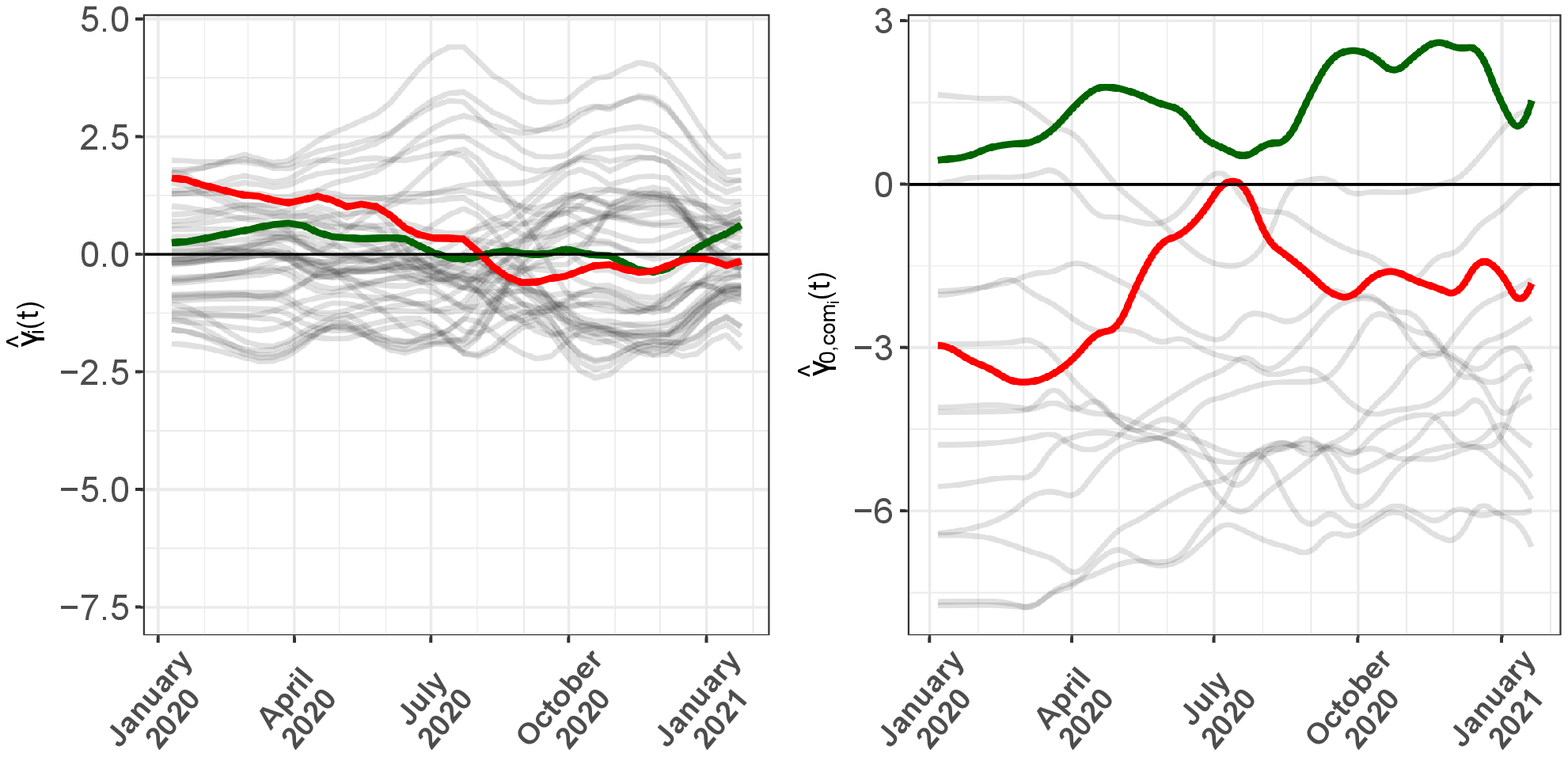}}
    \caption{Estimated functional random effects under the commonly shared border based neighbourhood definition. Left:  spatially correlated functional random intercepts per province (Markov Random Field specification), with  estimated curves for the provinces of Madrid (green) and Lleida (red) highlighted, computed from 47 continental provinces. Right: spatially uncorrelated functional random intercepts per  community, with  estimated curves for the communities of Madrid (green) and Catalonia (red) highlighted, computed from 15 communities.}
    \label{fig:mrfeffectqueen}
\end{figure}

Comparing the estimated  spatially correlated and uncorrelated functional random intercepts of the commonly shared border based model (Figure \ref{fig:mrfeffectqueen}) with the random effect terms under the Gabriel graph specification shows some differences, also in sign. 
We note that the Gabriel and the commonly shared border approaches yield different neighbourhood specifications and different numbers of regional neighbours, e.g. Lleida has 5 neighbours under the commonly shared border approaches compared to 3 under the Gabriel graph specification, such that some differences in spatial smoothing are not unexpected.  

\begin{figure}[h!]
    \centering
    \makebox{
    \includegraphics[scale=0.52]{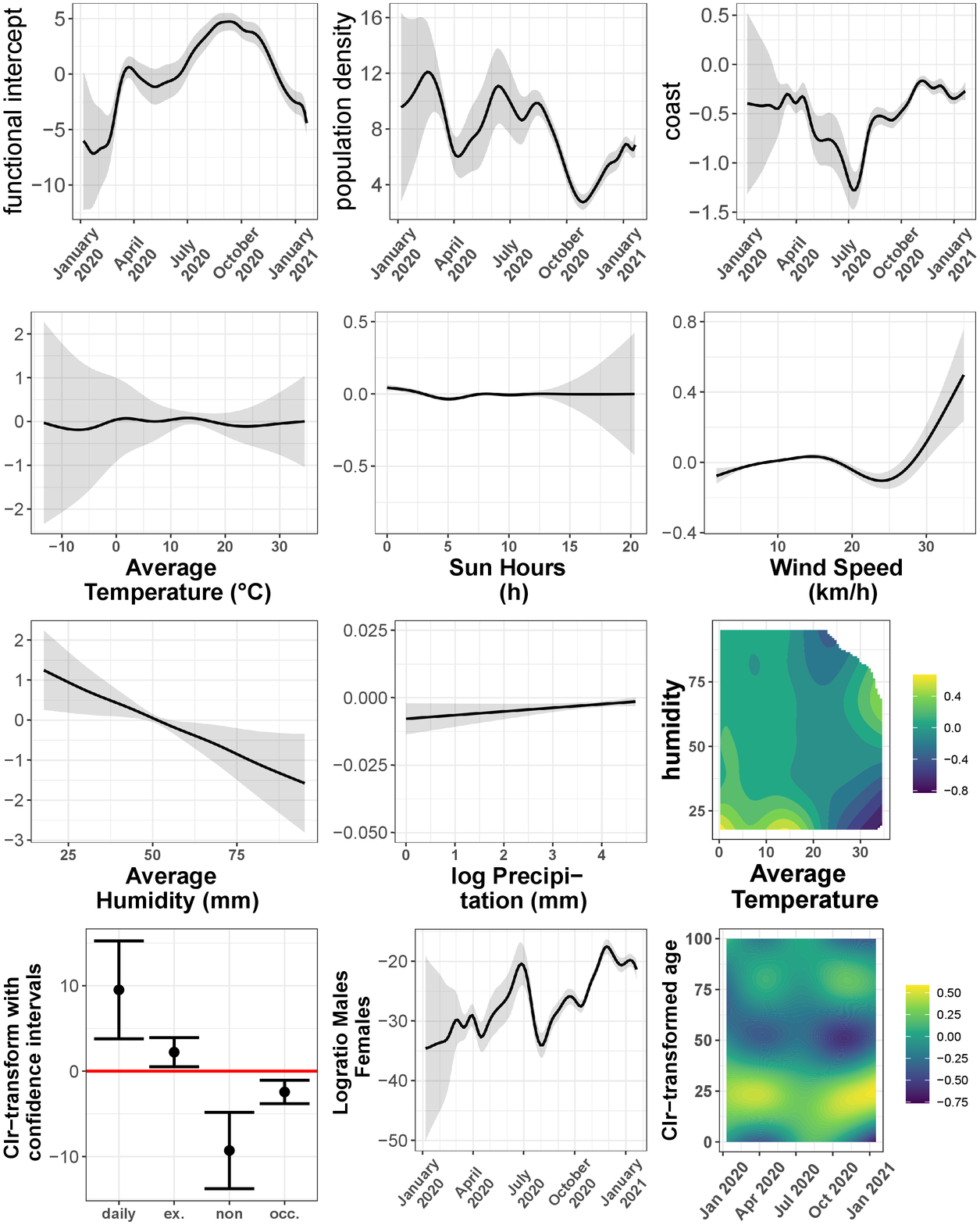}}
    \caption{Functional intercept and effects under the Gabriel graph neighbourhood specification for continental Spain.}
    \label{fig:m1_effect_nonline_gabriel_continental}
\end{figure}

To better understand the variation in the effect   between the Gabriel graph and the commonly shared boarder based models, we additionally fitted a Gabriel graph model restricted to continental Spain only. This model explains  $96.4\%$ of the deviance (scale = 14.288). Table \ref{tab:m1:gabriel_continental} and Figure \ref{fig:m1_effect_nonline_gabriel_continental} give the corresponding results. Results are similar to those for the commonly shared border neighbourhood specification, reinforcing the conclusion that differences seen there to results in the main paper are due to the reduced data set rather than the specified neighbourhood structure. Besides the same differences in the main effects for temperature and humidity compared to the main results (while the interaction stays similar), we mainly observe some differences in the size while not general trend of the estimated smoking effect.  
\begin{table}
\caption{Constant coefficients of the functional generalised additive model under the Gabriel graph neighbourhood specification for continental Spain \label{tab:m1:gabriel_continental}}
\centering
\fbox{%
\begin{tabular}{l*{5}{c}}
 & \em     $\beta$ &\em $\exp(\beta)$ &\em s.e. &\em $t$-value & Pr$(>|t|)$\\
       \hline
Intercept & 5.79 & 326.56 & 2.30 & 2.52 & 0.01 \\ 
Lockdown 1& -0.13 & 0.88 & 0.04 & -3.06 & 0.00 \\ 
Lockdown 2 & -0.03 & 0.97 & 0.14 & -0.21 & 0.84 \\ 
Lockdown 3 & 0.03 & 1.03 & 0.13 & 0.23 & 0.81 \\ 
ilr(smoke 1) & -8.45 & 0.00 & 2.32 & -3.64 & 0.00 \\ 
ilr(smoke 2) & -1.08 & 0.34 & 1.63 & -0.66 & 0.51 \\ 
ilr(smoke 3) & -10.74 & 0.00 & 2.63 & -4.08 & 0.00 \\ 
Monday & 0.35 & 1.42 & 0.01 & 34.60 & 0.00 \\ 
Tuesday & 0.42 & 1.52 & 0.01 & 42.10 & 0.00 \\ 
Wednesday & 0.38 & 1.46 & 0.01 & 37.11 & 0.00 \\ 
Thursday & 0.33 & 1.39 & 0.01 & 32.70 & 0.00 \\ 
Friday & 0.39 & 1.48 & 0.01 & 39.23 & 0.00 \\ 
Saturday & 0.14 & 1.15 & 0.01 & 13.26 & 0.00 \\ 
Transport & 0.60 & 1.83 & 0.01 & 39.18 & 0.00 \\ 
GDP & -0.32 & 0.73 & 0.02 & -13.13 & 0.00 \\ 
Rain & 0.03 & 1.03 & 0.01 & 3.54 & 0.00 \\ 
   \hline
\end{tabular}}
\end{table}

\begin{figure}[h!]
\centering
\makebox{
    \includegraphics[scale=0.5]{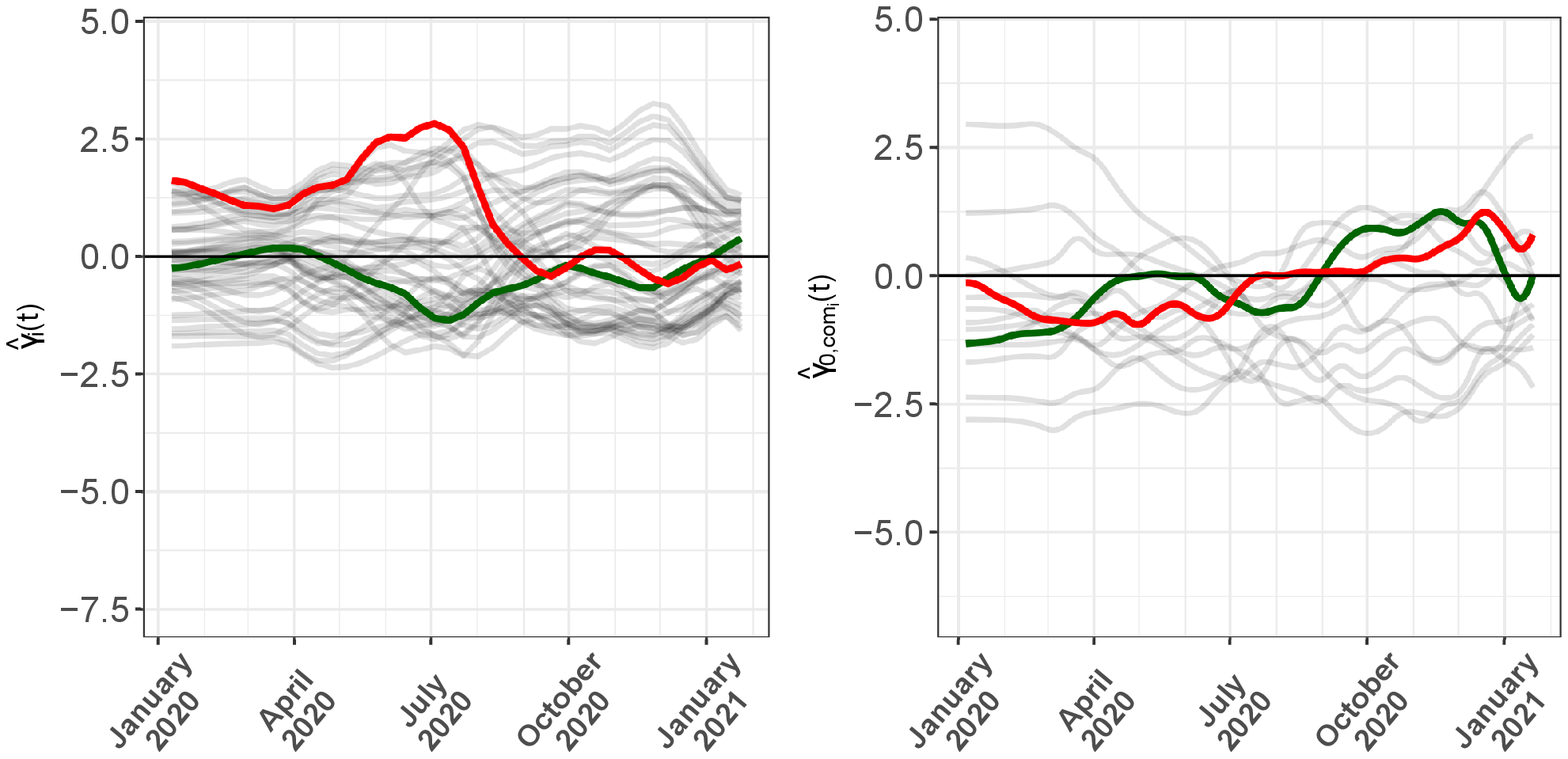}}
    \caption{Estimated functional random effects under the Gabriel graph based neighbourhood definition computed for continental Spain only. Left:  spatially correlated functional random intercepts per province (Markov Random Field specification), with  estimated curves for the provinces of Madrid (green) and Lleida (red) highlighted, computed from 47 continental provinces. Right: spatially uncorrelated functional random intercepts per  community, with  estimated curves for the communities of Madrid (green) and Catalonia (red) highlighted, computed from 15 communities.}
    \label{fig:mrfeffect_gabriel_continental}
\end{figure}

\subsection{Lag Effects}

To account for the uncertainty in the lag-specification for the weather covariates (given the imperfect reconstruction of symptom onset and variation in the incubation period), we constructed identical models to the one in the main paper using a 4-day, 6-day, 7-day, and 8-day lag for the concurrent weather effects and interaction surface. In all models, we specified the spatial neighbourhood structure of the MRF through a Gabriel graph approach. For all models, we found only small differences of the estimated scale parameters (with the smallest scale parameter under the 5-day lag specification) while all models (slightly lower for the 4-day lag model) explained $96\%$ of the deviance (see Table \ref{tab:gof:lags}). 
\begin{table}
\caption{Deviance explained and scale parameter for different lag-specifications of the concurrent weather effects\label{tab:gof:lags}}
\centering
\fbox{%
\begin{tabular}{lcc}
 \text{lag}& \text{deviance explained} $(\%)$ &\text{scale} \\
       \hline
4-day lag & 95.7& 16.044\\  
5-day lag & 96 &  15.099 \\
6-day lag & 96 & 15.271\\
7-day lag & 96 & 15.337\\
8-day lag & 96 & 15.219\\
   \hline
\end{tabular}}
\end{table}
The constant coefficients for all alternative lag specifications are reported in Tables \ref{tab:m1:minus1} to \ref{tab:m1:plus3}. 
Results are overall similar to the main model in effect size and significance, however with the effect of the first lockdown indicator becoming smaller and non-significant for 4-day, 6-day and 7-day lag models. The largest difference in effects is seen in the weekday effects of the 4-day lag model, where positive effects of weekdays compared to Sunday become smaller and even slightly negative for Wednesday and Sunday. The other effects either stay roughly constant (e.g. transport or GDP) or change somewhat in size (e.g. rain, between 0.02 and 0.07), but not in sign or significance. Overall, effects are relatively stable across lag specifications. 
\begin{table}
\caption{Constant coefficients of the functional generalised additive model under the Gabriel graph based neighbourhood definition for the 4-day lag model \label{tab:m1:minus1}}
\centering
\fbox{%
\begin{tabular}{l*{5}{c}}
 & \em     $\beta$ &\em $\exp(\beta)$ &\em s.e. &\em $t$-value & Pr$(|T|>|t|)$\\
       \hline
(Intercept) & 1.99 & 7.32 & 1.45 & 1.37 & 0.17 \\ 
  Lockdown 1& -0.08 & 0.93 & 0.04 & -1.74 & 0.08 \\ 
  Lockdown 2 & 0.01 & 1.01 & 0.14 & 0.07 & 0.94 \\ 
  Lockdown 3 & 0.03 & 1.03 & 0.14 & 0.25 & 0.80 \\ 
  ilr(smoke 1) & -5.51 & 0.00 & 1.98 & -2.78 & 0.00 \\ 
  ilr(smoke 2)& -0.99 & 0.37 & 1.41 & -0.70 & 0.48 \\ 
  ilr(smoke 3) & -5.18 & 0.01 & 2.25 & -2.30 & 0.02 \\ 
  Monday & 0.14 & 1.15 & 0.01 & 16.94 & 0.00 \\ 
  Tuesday & 0.20 & 1.22 & 0.01 & 24.18 & 0.00 \\ 
  Wednesday & -0.01 & 0.99 & 0.01 & -1.99 & 0.05 \\ 
  Thursday & 0.12 & 1.12 & 0.01 & 13.65 & 0.00 \\ 
  Friday & 0.18 & 1.20 & 0.01 & 21.69 & 0.00 \\ 
  Saturday & -0.06 & 0.94 & 0.01 & -6.77 & 0.00 \\ 
  Transport & 0.47 & 1.61 & 0.02 & 30.43 & 0.00 \\ 
  GDP & -0.39 & 0.68 & 0.02 & -15.48 & 0.00 \\ 
  Rain & 0.02 & 1.02 & 0.01 & 2.74 & 0.01 \\ 
   \hline
\end{tabular}}
\end{table}

\begin{table}
\caption{Constant coefficients of the functional generalised additive model under the Gabriel graph based neighbourhood definition for the 6-day lag model \label{tab:m1:plus1}}
\centering
\fbox{%
\begin{tabular}{l*{5}{c}}
 & \em     $\beta$ &\em $\exp(\beta)$ &\em s.e. &\em $t$-value & Pr$(|T|>|t|)$\\
       \hline
Intercept & 3.20 & 24.55 & 1.30 & 2.46 & 0.01 \\ 
Lockdown 1 & -0.08 & 0.92 & 0.04 & -1.83 & 0.07 \\ 
Lockdown 2 & 0.02 & 1.02 & 0.14 & 0.15 & 0.88 \\ 
Lockdown 3 & 0.08 & 1.08 & 0.13 & 0.61 & 0.54 \\ 
ilr(smoke 1) & -5.22 & 0.00 & 1.75 & -2.99 & 0.00 \\ 
ilr(smoke 2) & -1.10 & 0.33 & 1.25 & -0.88 & 0.38 \\ 
ilr(smoke 3) & -5.77 & 0.00 & 1.98 & -2.91 & 0.00 \\ 
Monday & 0.34 & 1.40 & 0.01 & 33.60 & 0.00 \\ 
Tuesday & 0.41 & 1.51 & 0.01 & 40.56 & 0.00 \\ 
Wednesday & 0.38 & 1.46 & 0.01 & 36.59 & 0.00 \\ 
Thursday & 0.33 & 1.39 & 0.01 & 31.71 & 0.00 \\ 
Friday & 0.38 & 1.47 & 0.01 & 38.08 & 0.00 \\ 
Saturday & 0.13 & 1.14 & 0.01 & 12.48 & 0.00 \\ 
Transport & 0.46 & 1.59 & 0.01 & 30.43 & 0.00 \\ 
GDP & -0.39 & 0.68 & 0.02 & -15.87 & 0.00 \\ 
Rain & 0.04 & 1.04 & 0.01 & 3.85 & 0.00 \\
\hline
\end{tabular}}
\end{table}

\begin{table}
\caption{Constant coefficients of the functional generalised additive model under the Gabriel graph based neighbourhood definition for the 7-day lag model \label{tab:m1:plus2}}
\centering
\fbox{%
\begin{tabular}{l*{5}{c}}
 & \em     $\beta$ &\em $\exp(\beta)$ &\em s.e. &\em $t$-value & Pr$(|T|>|t|)$\\
       \hline
Intercept & 1.30 & 3.66 & 1.49 & 0.87 & 0.38 \\ 
Lockdown 1 & -0.02 & 0.98 & 0.04 & -0.46 & 0.64 \\ 
Lockdown 2 & 0.15 & 1.17 & 0.14 & 1.08 & 0.28 \\ 
Lockdown 3 & 0.15 & 1.16 & 0.12 & 1.22 & 0.22 \\ 
ilr(smoke 1) & -5.63 & 0.00 & 2.13 & -2.65 & 0.01 \\ 
ilr(smoke 2) & -0.97 & 0.38 & 1.51 & -0.64 & 0.52 \\ 
ilr(smoke 3) & -5.06 & 0.01 & 2.41 & -2.10 & 0.04 \\ 
Monday & 0.34 & 1.41 & 0.01 & 33.95 & 0.00 \\ 
Tuesday & 0.42 & 1.52 & 0.01 & 41.80 & 0.00 \\ 
Wednesday & 0.38 & 1.47 & 0.01 & 36.96 & 0.00 \\ 
Thursday & 0.33 & 1.39 & 0.01 & 32.22 & 0.00 \\ 
Friday & 0.39 & 1.48 & 0.01 & 38.35 & 0.00 \\ 
Saturday & 0.14 & 1.15 & 0.01 & 13.50 & 0.00 \\ 
Transport & 0.47 & 1.60 & 0.01 & 30.75 & 0.00 \\ 
GDP & -0.39 & 0.68 & 0.02 & -15.80 & 0.00 \\ 
Rain & 0.05 & 1.05 & 0.01 & 4.12 & 0.00 \\
\hline
\end{tabular}}
\end{table}

\begin{table}
\caption{Constant coefficients of the functional generalised additive model under the Gabriel graph based neighbourhood definition for the 8-day lag model \label{tab:m1:plus3}}
\centering
\fbox{%
\begin{tabular}{l*{5}{c}}
 & \em     $\beta$ &\em $\exp(\beta)$ &\em s.e. &\em $t$-value & Pr$(|T|>|t|)$\\
       \hline
Intercept & 1.33 & 3.79 & 1.48 & 0.90 & 0.37 \\ 
Lockdown 1 & -0.10 & 0.91 & 0.04 & -2.20 & 0.03 \\ 
Lockdown 2 & 0.13 & 1.14 & 0.14 & 0.95 & 0.34 \\ 
Lockdown 3 & 0.14 & 1.15 & 0.12 & 1.16 & 0.24 \\ 
ilr(smoke 1) & -5.57 & 0.00 & 2.10 & -2.65 & 0.01 \\ 
ilr(smoke 2) & -0.96 & 0.38 & 1.50 & -0.64 & 0.52 \\ 
ilr(smoke 3) & -5.11 & 0.01 & 2.38 & -2.14 & 0.03 \\ 
Monday & 0.34 & 1.40 & 0.01 & 33.44 & 0.00 \\ 
Tuesday & 0.41 & 1.51 & 0.01 & 41.15 & 0.00 \\ 
Wednesday & 0.38 & 1.46 & 0.01 & 37.16 & 0.00 \\ 
Thursday & 0.34 & 1.40 & 0.01 & 32.79 & 0.00 \\ 
Friday & 0.38 & 1.47 & 0.01 & 37.98 & 0.00 \\ 
Saturday & 0.14 & 1.14 & 0.01 & 12.79 & 0.00 \\ 
Transport & 0.46 & 1.59 & 0.01 & 30.44 & 0.00 \\ 
GDP & -0.38 & 0.68 & 0.02 & -15.65 & 0.00 \\ 
Rain & 0.07 & 1.07 & 0.01 & 6.60 & 0.00 \\ 
\hline
\end{tabular}}
\end{table}

\begin{figure}[h!]
    \centering
    \makebox{
    \includegraphics[scale=0.52]{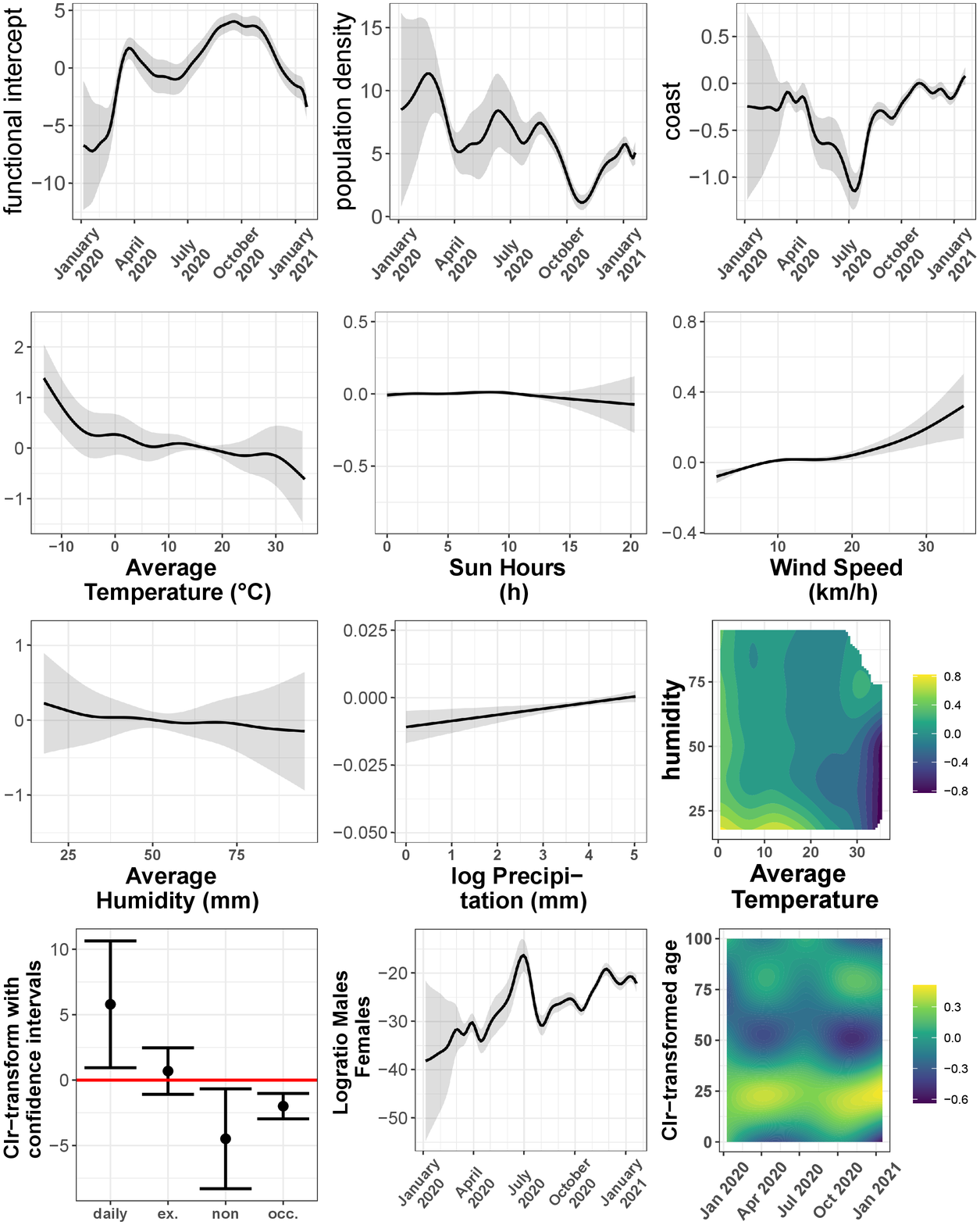}}
\caption{Functional intercept and effects under the 4-day lag specification of the weather covariates.}
\label{fig:minus1lag}
\end{figure}

\begin{figure}[h!]
    \centering
    \makebox{
    \includegraphics[scale=0.52]{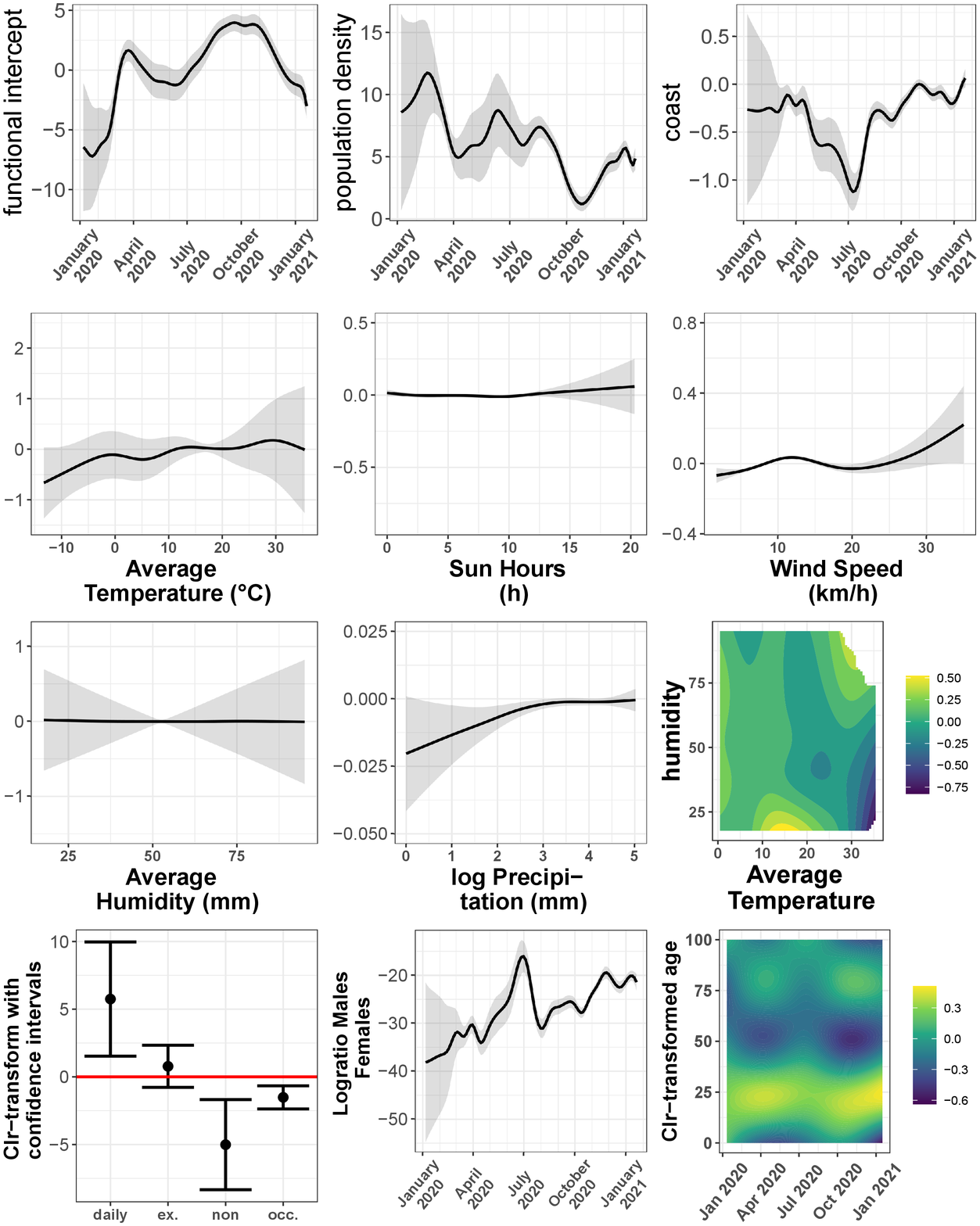}}
\caption{Functional intercept and effects under the 6-day lag specification of the weather covariates.}
\label{fig:plus1lag}
\end{figure}

\begin{figure}[h!]
    \centering
    \makebox{
    \includegraphics[scale=0.52]{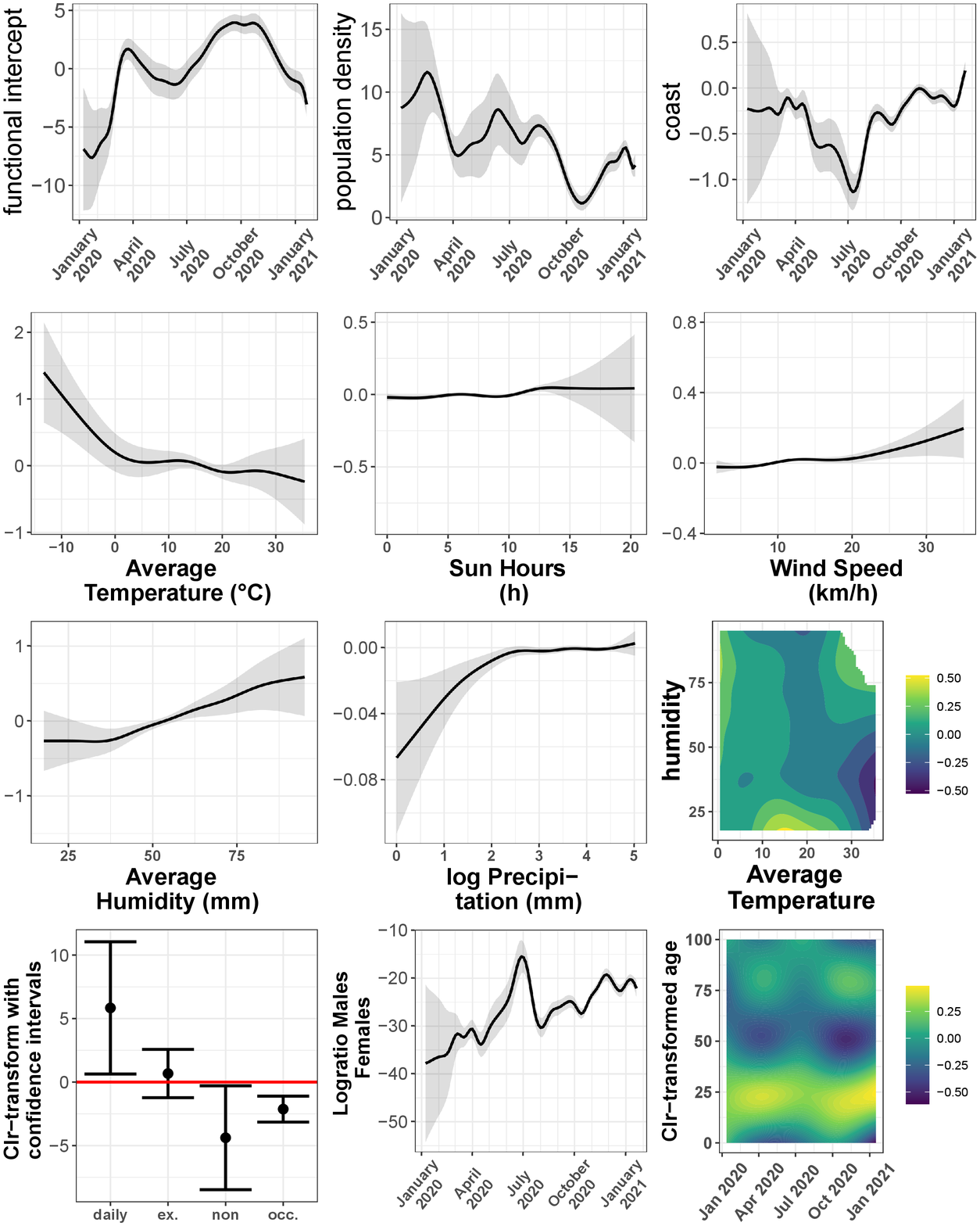}}
\caption{Functional intercept and effects under the 7-day lag specification of the weather covariates.}
\label{fig:plus2lag}
\end{figure}

\begin{figure}[h!]
    \centering
    \makebox{
    \includegraphics[scale=0.52]{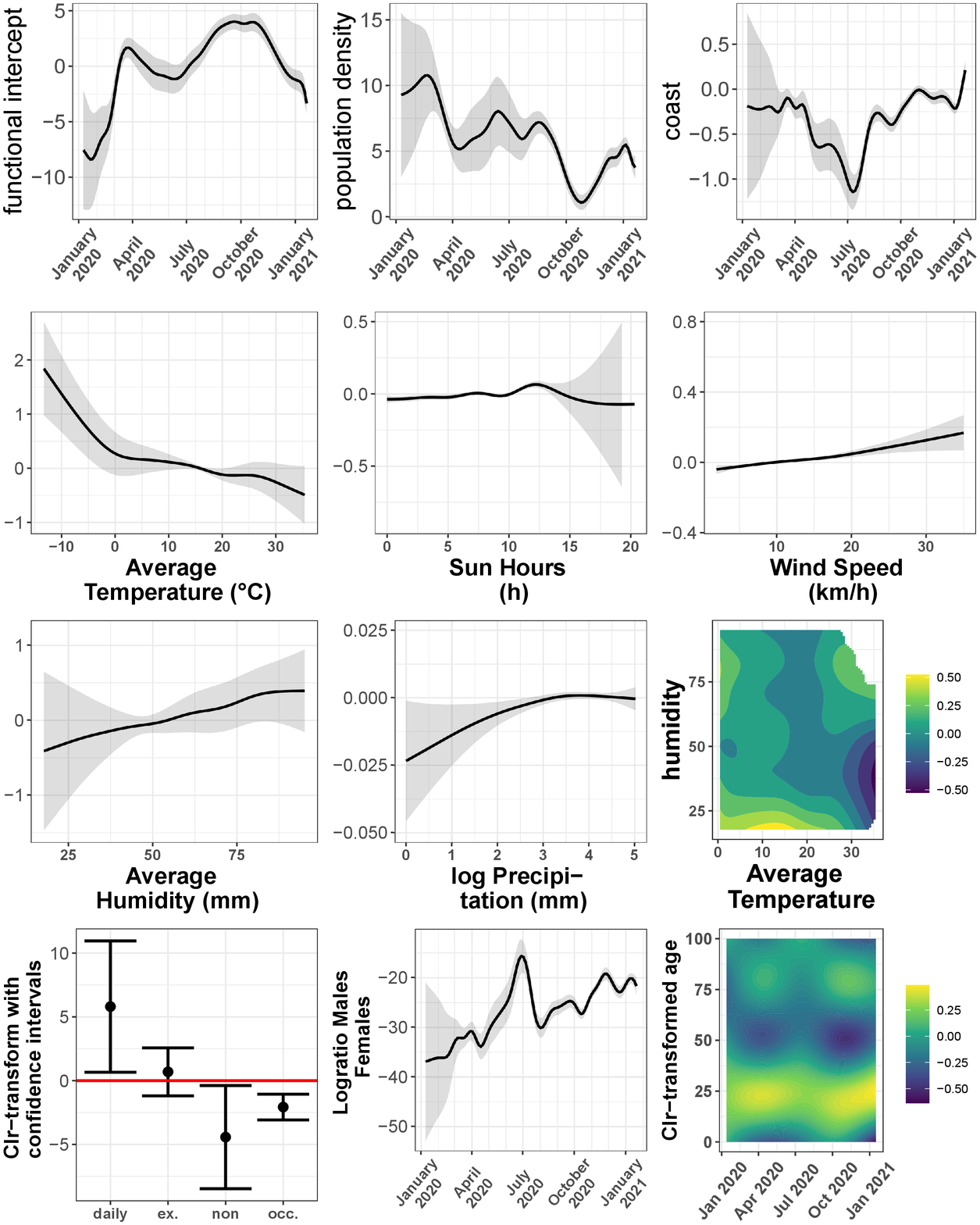}}
\caption{Functional intercept and effects under the 8-day lag specification of the weather covariates.}
\label{fig:plus3lag}
\end{figure}

All four lag models support the reported structure of the functional intercept and smoothly time-varying effects of the population density and coast and the effects of the compositional covariates  (i.e.\ all non-weather variables) (see Figures \ref{fig:minus1lag} to \ref{fig:plus3lag}). For the non-linear effects and interaction effect of the lagged weather variables themselves, general trends are mostly similar to the 5-day lag model, although effect function  shapes are not always identical and effects tend on average to be somewhat smaller than for the 5-day model. However, for the non-linear effects of the lagged average temperature and lagged humidity we observe some variation under the 6-day lag specification, with humidity having no effect and the trend for average temperature reverting, while the corresponding confidence band completely contains the zero line. The main effects plus interaction surface, however, shows similar overall trends with smaller values for both low temperature and low humidity or both high temperature and high humidity. Overall, the observed patterns are consistent with our hypothesized 5-day lag best capturing the time lag between observed incidence and infection, with results mostly not being  sensitive to that choice.  

\begin{figure}[h!]
\centering
\makebox{
    \includegraphics[scale=0.5]{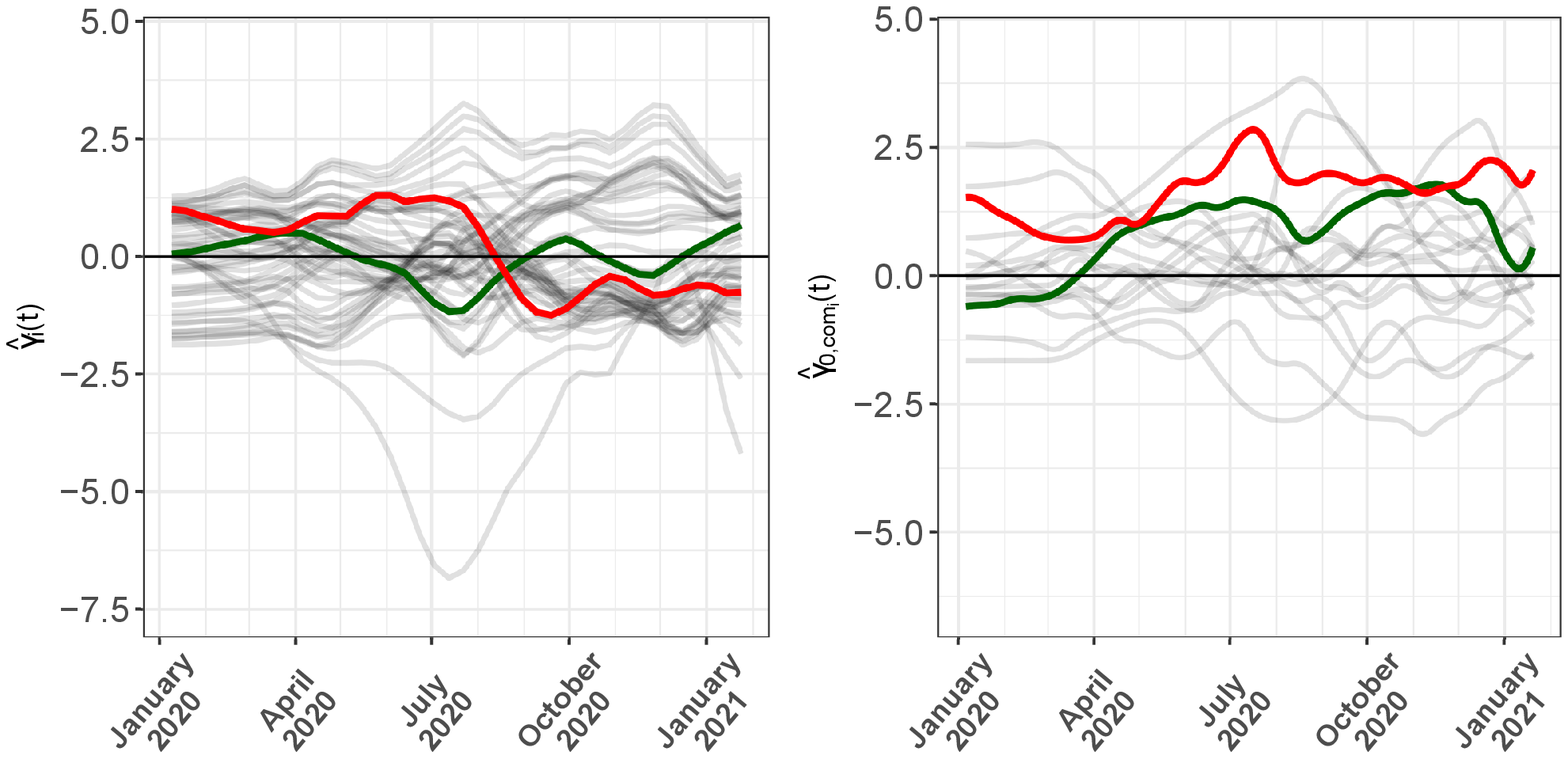}}
    \caption{Estimated functional random effects under the Gabriel graph based neighbourhood definition with time-varying weather effects under the 4-day lag model. Left:  spatially correlated functional random intercepts per region (Markov Random Field specification), with  estimated curves for the provinces of Madrid (green) and Lleida (red) highlighted. Right: spatially uncorrelated functional random intercepts per  community, with  estimated curves for the communities of Madrid (green) and Catalonia (red) highlighted.}
    \label{fig:mrf_effect_4day}
\end{figure}

\begin{figure}[h!]
\centering
\makebox{
    \includegraphics[scale=0.5]{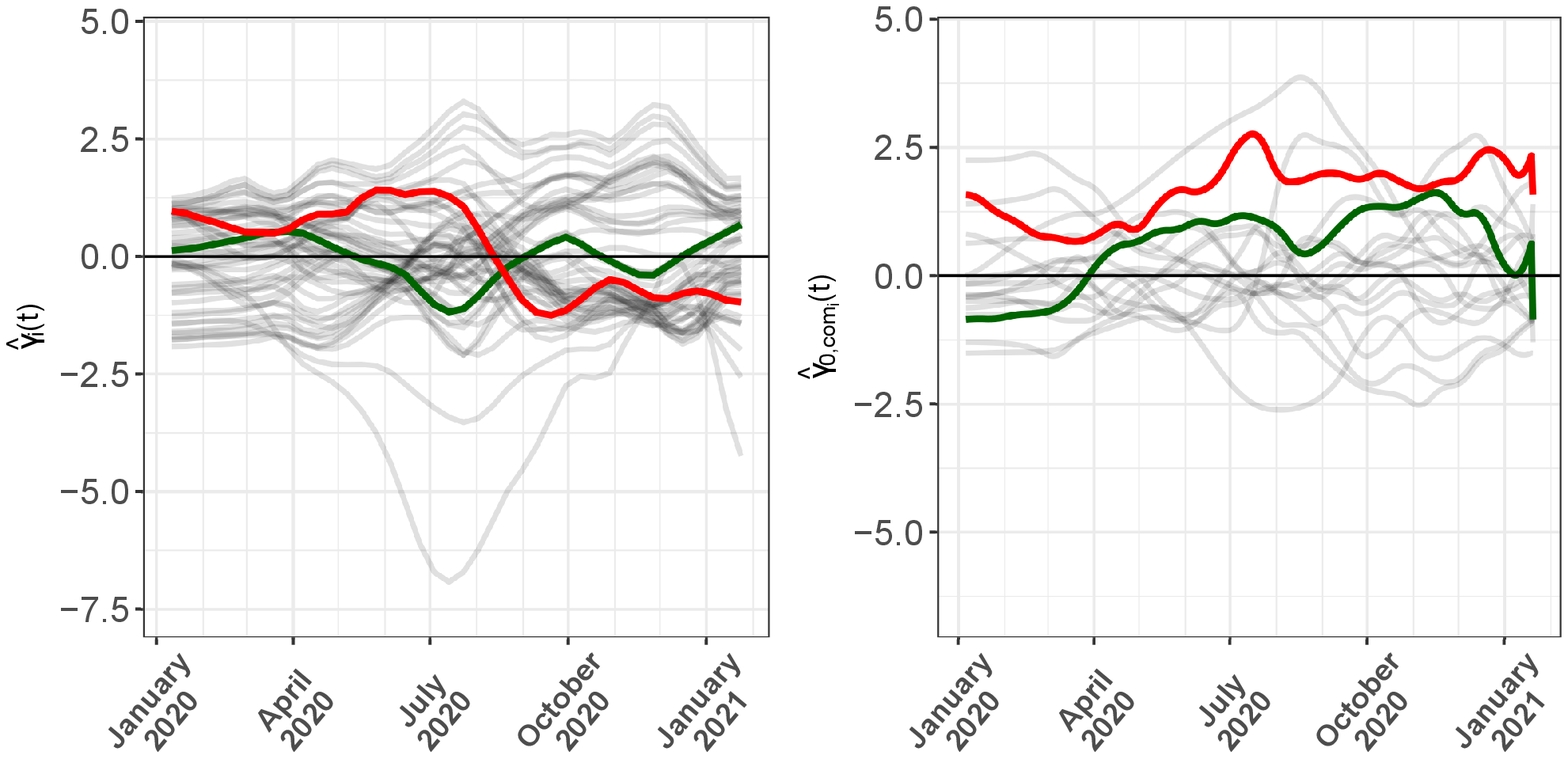}}
    \caption{Estimated functional random effects under the Gabriel graph based neighbourhood definition with time-varying weather effects under the 6-day lag model. Left: spatially correlated functional random intercepts (Markov Random Field specification), with  estimated curves for the provinces of Madrid (green) and Lleida (red) highlighted. Right: spatially uncorrelated functional random intercepts per  community, with  estimated curves for the communities of Madrid (green) and Catalonia (red) highlighted.}
    \label{fig:plus1_effect_mrf}
\end{figure}

\begin{figure}[h!]
\centering
\makebox{
    \includegraphics[scale=0.5]{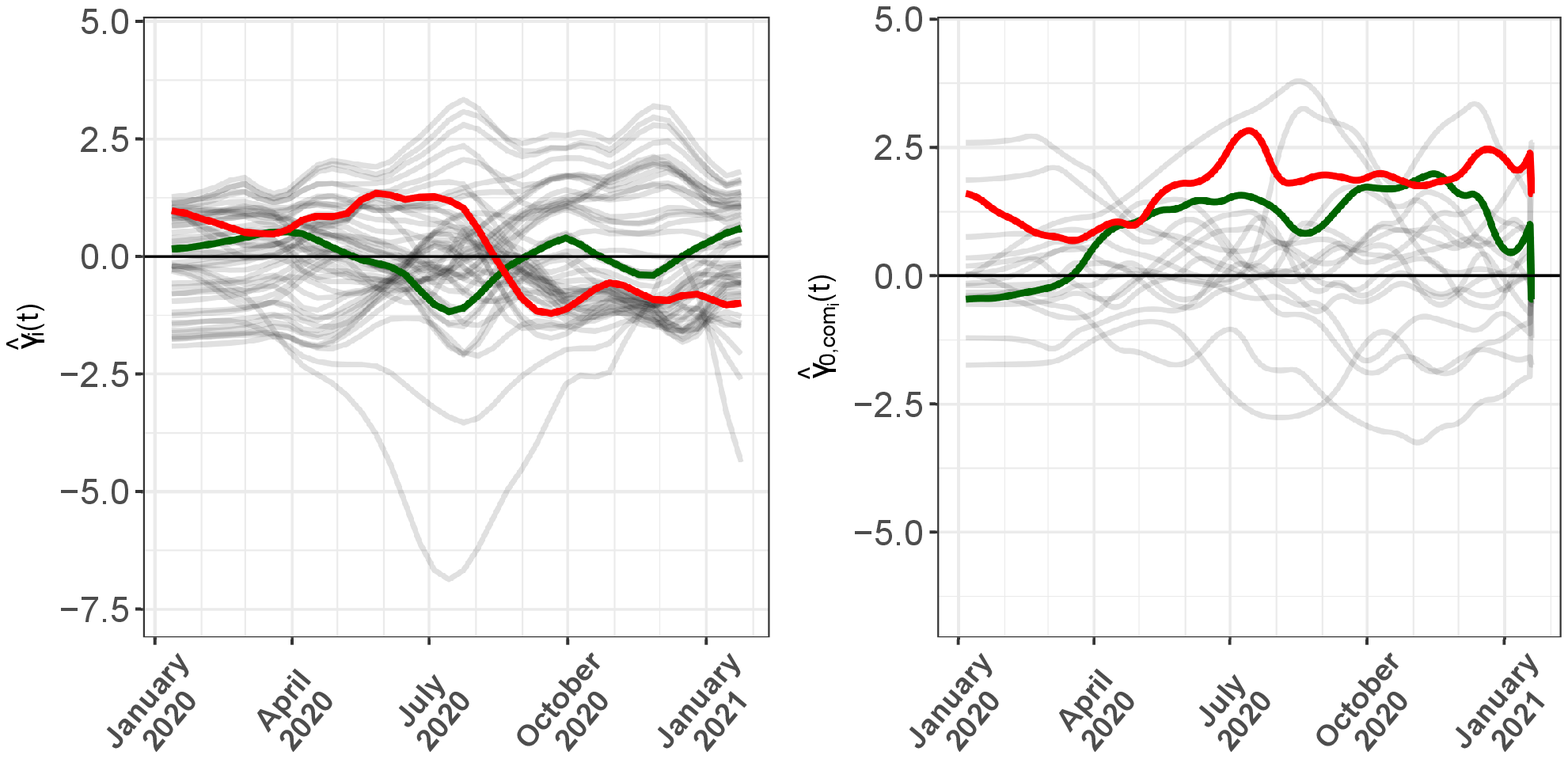}}
    \caption{Estimated functional random effects under the Gabriel graph based neighbourhood definition with time-varying weather effects under the 7-day lag model. Left: spatially correlated functional random intercepts (Markov Random Field specification), with  estimated curves for the provinces of Madrid (green) and Lleida (red) highlighted. Right: spatially uncorrelated functional random intercepts per  community, with  estimated curves for the communities of Madrid (green) and Catalonia (red) highlighted.}
    \label{fig:plus2_effect_mrf}
\end{figure}

\begin{figure}[h!]
\centering
\makebox{
    \includegraphics[scale=0.5]{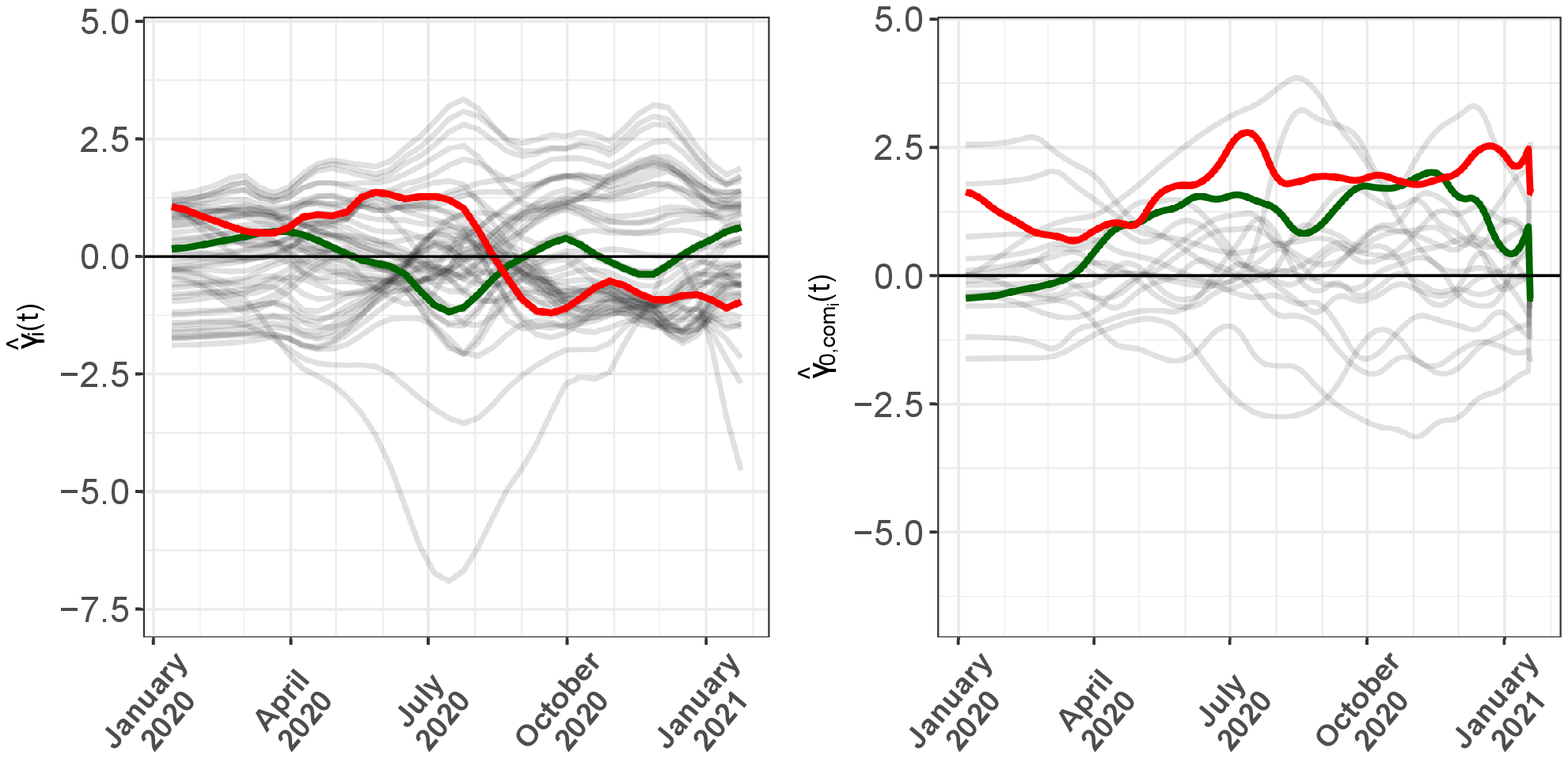}}
    \caption{Estimated functional random effects under the Gabriel graph based neighbourhood definition with time-varying weather effects under the 8-day lag model. Left: spatially correlated functional random intercepts (Markov Random Field specification), with  estimated curves for the provinces of Madrid (green) and Lleida (red) highlighted. Right: spatially uncorrelated functional random intercepts per  community, with  estimated curves for the communities of Madrid (green) and Catalonia (red) highlighted.}
    \label{fig:plus3_effect_mrf}
\end{figure}

Finally, all four different lag-specifications barely affect the  spatially correlated or the spatially uncorrelated functional random intercepts depicted in Figures \ref{fig:mrf_effect_4day} to \ref{fig:plus3_effect_mrf}.

\bibliographystyle{rss}
\bibliography{fda}

\end{document}